\documentclass[preprint]{aastex63}

\revised{April 22, 2022} 
\submitjournal{AJ}

\usepackage{lineno}

\shorttitle{IXPE Modulation Factor}
\shortauthors{Di Marco et al.}
\DeclareOldFontCommand{\bf}{\normalfont\bfseries}{\mathbf} 
\providecommand{\DIFadd}[1]{{\bf #1}} 
\providecommand{\DIFdel}[1]{} 
\providecommand{\DIFaddbegin}{} 
\providecommand{\DIFaddend}{} 
\providecommand{\DIFdelbegin}{} 
\providecommand{\DIFdelend}{} 
\RequirePackage{listings} 
\RequirePackage{color} 
\lstdefinelanguage{DIFcode}{ 
  moredelim=[il][\color{white}\tiny]{\%DIF\ <\ }, 
  moredelim=[il][\sffamily\bfseries]{\%DIF\ >\ } 
} 
\lstdefinestyle{DIFverbatimstyle}{ 
	language=DIFcode, 
	basicstyle=\ttfamily, 
	columns=fullflexible, 
	keepspaces=true 
} 
\lstnewenvironment{DIFverbatim}{\lstset{style=DIFverbatimstyle}}{} 
\lstnewenvironment{DIFverbatim*}{\lstset{style=DIFverbatimstyle,showspaces=true}}{} 

\begin{document}

\title{Calibration of \DIFdelbegin \DIFdel{modulation factor and of }\DIFdelend the \DIFdelbegin \DIFdel{polarization angle of }\DIFdelend IXPE focal plane X-ray polarimeters \DIFaddbegin \DIFadd{to polarized radiation}\DIFaddend }

\correspondingauthor{Alessandro Di Marco}
\email{alessandro.dimarco@inaf.it}

\author[0000-0003-0331-3259]{Alessandro Di Marco}
\affiliation{INAF -- IAPS, via Fosso del Cavaliere, 100, Rome, Italy I-00133}

\author[0000-0003-1533-0283]{Sergio Fabiani}
\affiliation{INAF -- IAPS, via Fosso del Cavaliere, 100, Rome, Italy I-00133}

\author[0000-0001-8916-4156]{Fabio La Monaca}
\affiliation{INAF -- IAPS, via Fosso del Cavaliere, 100, Rome, Italy I-00133}

\author[0000-0003-3331-3794]{Fabio Muleri}
\affiliation{INAF -- IAPS, via Fosso del Cavaliere, 100, Rome, Italy I-00133}

\author[0000-0002-9774-0560]{John Rankin}
\affiliation{INAF -- IAPS, via Fosso del Cavaliere, 100, Rome, Italy I-00133}
\affiliation{Università di Roma “La Sapienza”, Dipartimento di Fisica, Piazzale Aldo Moro 2, Rome, Italy I-00185}
\affiliation{Università di Roma "Tor Vergata", Dipartimento di Fisica', Via della Ricerca Scientifica, 1, I-00133 Roma, Italy}

\author[0000-0002-7781-4104]{Paolo Soffitta}
\affiliation{INAF -- IAPS, via Fosso del Cavaliere, 100, Rome, Italy I-00133}

\author{Fei Xie}
\affiliation{INAF -- IAPS, via Fosso del Cavaliere, 100, Rome, Italy I-00133}

\author{Fabrizio Amici}
\affiliation{INAF -- IAPS, via Fosso del Cavaliere, 100, Rome, Italy I-00133}

\author{Primo attinà}
\affiliation{INAF/Osservatorio Astrofisico di Torino, Via Osservatorio 20, I-10025 Pino Torinese (TO), Italy}

\author{Matteo Bachetti}
\affiliation{INAF-OAC, Via della Scienza 5, I-09047 Selargius (CA), Italy}

\author{Luca Baldini}
\affiliation{fUniversit`a di Pisa, Dipartimento di Fisica Enrico Fermi, Largo B. Pontecorvo 3, I-56127 Pisa, Italy}
\affiliation{INFN-Pisa, Largo B. Pontecorvo 3, I-56127 Pisa, Italy}

\author{Mattia Barbanera}
\affiliation{INFN-Pisa, Largo B. Pontecorvo 3, I-56127 Pisa, Italy}
\affiliation{Universit`a di Pisa, Dipartimento di Ingegneria 
dell’Informazione, Via G. Caruso 16, I-56122 Pisa, Italy}

\author{Wayne Baumgartner}
\affiliation{NASA Marshall Space Flight Center, Huntsville, AL 35812, USA}

\author{Ronaldo Bellazzini}
\affiliation{INFN-Pisa, Largo B. Pontecorvo 3, I-56127 Pisa, Italy}

\author{Fabio Borotto}
\affiliation{INFN-Torino, Via P. Giuria, 1, I-10125 Torino, Italy}

\author{Alessandro Brez}
\affiliation{INFN-Pisa, Largo B. Pontecorvo 3, I-56127 Pisa, Italy}

\author{Daniele Brienza}
\affiliation{INAF -- IAPS, via Fosso del Cavaliere, 100, Rome, Italy I-00133}

\author{Ciro Caporale}
\affiliation{INFN-Torino, Via P. Giuria, 1, I-10125 Torino, Italy}

\author{Claudia Cardelli}
\affiliation{INFN-Pisa, Largo B. Pontecorvo 3, I-56127 Pisa, Italy}

\author{Rita Carpentiero}
\affiliation{ASI, Via del Politecnico snc, I-00133 Roma, Italy}

\author{Simone Castellano}
\affiliation{INFN-Pisa, Largo B. Pontecorvo 3, I-56127 Pisa, Italy}

\author{Marco Castronuovo}
\affiliation{ASI, Via del Politecnico snc, I-00133 Roma, Italy}

\author{Luca Cavalli}
\affiliation{OHB Italia, Via Gallarate 150, I-20151 Milano, Italy}

\author{Elisabetta Cavazzuti}
\affiliation{ASI, Via del Politecnico snc, I-00133 Roma, Italy}

\author{Marco Ceccanti}
\affiliation{INFN-Pisa, Largo B. Pontecorvo 3, I-56127 Pisa, Italy}

\author{Mauro Centrone}
\affiliation{INAF/OAR, Via Frascati 33, I-00040, Monte Porzio Catone (RM)}

\author{Saverio Citraro}
\affiliation{INFN-Pisa, Largo B. Pontecorvo 3, I-56127 Pisa, Italy}

\author[0000-0003-4925-8523]{Enrico Costa}
\affiliation{INAF -- IAPS, via Fosso del Cavaliere, 100, Rome, Italy I-00133}

\author{Elisa D'Alba}
\affiliation{OHB Italia, Via Gallarate 150, I-20151 Milano, Italy}

\author{Fabio D'Amico}
\affiliation{ASI, Via del Politecnico snc, I-00133 Roma, Italy}

\author{Ettore Del Monte}
\affiliation{INAF -- IAPS, via Fosso del Cavaliere, 100, Rome, Italy I-00133}

\author{Sergio {Di Cosimo}}
\affiliation{INAF -- IAPS, via Fosso del Cavaliere, 100, Rome, Italy I-00133}

\author{Niccolò Di Lalla}
\affiliation{Kavli Institute, Department of Physics and SLAC, Stanford University, Stanford, CA 94305, USA}

\author{Giuseppe {Di Persio}}
\affiliation{INAF -- IAPS, via Fosso del Cavaliere, 100, Rome, Italy I-00133}

\author{Immacolata Donnarumma}
\affiliation{ASI, Via del Politecnico snc, I-00133 Roma, Italy}

\author{Yuri Evangelista}
\affiliation{INAF -- IAPS, via Fosso del Cavaliere, 100, Rome, Italy I-00133}

\author{Riccardo Ferrazzoli}
\affiliation{INAF -- IAPS, via Fosso del Cavaliere, 100, Rome, Italy I-00133}
\affiliation{Università di Roma “La Sapienza”, Dipartimento di Fisica, Piazzale Aldo Moro 2, Rome, Italy I-00185}
\affiliation{Università di Roma "Tor Vergata", Dipartimento di Fisica', Via della Ricerca Scientifica, 1, I-00133 Roma, Italy}

\author{Luca Latronico}
\affiliation{INFN-Torino, Via P. Giuria, 1, I-10125 Torino, Italy}

\author{Carlo Lefevre}
\affiliation{INAF -- IAPS, via Fosso del Cavaliere, 100, Rome, Italy I-00133}

\author{Pasqualino Loffredo}
\affiliation{INAF -- IAPS, via Fosso del Cavaliere, 100, Rome, Italy I-00133}

\author{Paolo Lorenzi}
\affiliation{OHB Italia, Via Gallarate 150, I-20151 Milano, Italy}

\author{Leonardo Lucchesi}
\affiliation{INFN-Pisa, Largo B. Pontecorvo 3, I-56127 Pisa, Italy}

\author{Carlo Magazzù}
\affiliation{INFN-Pisa, Largo B. Pontecorvo 3, I-56127 Pisa, Italy}

\author{Guido Magazzù}
\affiliation{INFN-Pisa, Largo B. Pontecorvo 3, I-56127 Pisa, Italy}

\author{Simone Maldera}
\affiliation{INFN-Torino, Via P. Giuria, 1, I-10125 Torino, Italy}

\author{Alberto Manfreda}
\affiliation{INFN-Pisa, Largo B. Pontecorvo 3, I-56127 Pisa, Italy}

\author{Elio Mangraviti}
\affiliation{OHB Italia, Via Gallarate 150, I-20151 Milano, Italy}

\author{Marco Marengo}
\affiliation{INFN-Torino, Via P. Giuria, 1, I-10125 Torino, Italy}

\author{Giorgio Matt}
\affiliation{Università Roma Tre, Dipartimento di Matematica e Fisica, Via della Vasca Navale 84, I-00146, Italy}

\author{Paolo Mereu}
\affiliation{INFN-Torino, Via P. Giuria, 1, I-10125 Torino, Italy}

\author{Massimo Minuti}
\affiliation{INFN-Pisa, Largo B. Pontecorvo 3, I-56127 Pisa, Italy}

\author{Alfredo Morbidini}
\affiliation{INAF -- IAPS, via Fosso del Cavaliere, 100, Rome, Italy I-00133}

\author{Federico Mosti}
\affiliation{INFN-Torino, Via P. Giuria, 1, I-10125 Torino, Italy}

\author{Hikmat Nasimi}
\affiliation{INFN-Pisa, Largo B. Pontecorvo 3, I-56127 Pisa, Italy}

\author{Barbara Negri}
\affiliation{ASI, Via del Politecnico snc, I-00133 Roma, Italy}

\author{Alessio Nuti}
\affiliation{INFN-Pisa, Largo B. Pontecorvo 3, I-56127 Pisa, Italy}

\author{Stephen L. O'Dell}
\affiliation{NASA Marshall Space Flight Center, Huntsville, AL 35812, USA}

\author{Leonardo Orsini}
\affiliation{INFN-Pisa, Largo B. Pontecorvo 3, I-56127 Pisa, Italy}

\author{Matteo Perri}
\affiliation{INAF/OAR, Via Frascati 33, I-00040, Monte Porzio Catone (RM)}

\author{Melissa Pesce-Rollins}
\affiliation{INFN-Pisa, Largo B. Pontecorvo 3, I-56127 Pisa, Italy}

\author{Raffaele Piazzolla}
\affiliation{ASI, Via del Politecnico snc, I-00133 Roma, Italy}

\author{Stefano Pieraccini}
\affiliation{OHB Italia, Via Gallarate 150, I-20151 Milano, Italy}

\author{Maura Pilia}
\affiliation{INAF-OAC, Via della Scienza 5, I-09047 Selargius (CA), Italy}

\author{Michele Pinchera}
\affiliation{INFN-Pisa, Largo B. Pontecorvo 3, I-56127 Pisa, Italy}

\author{Alessandro Profeti}
\affiliation{INFN-Pisa, Largo B. Pontecorvo 3, I-56127 Pisa, Italy}

\author{Simonetta Puccetti}
\affiliation{ASI, Via del Politecnico snc, I-00133 Roma, Italy}

\author{Brian D. Ramsey}
\affiliation{NASA Marshall Space Flight Center, Huntsville, AL 35812, USA}

\author{Ajay Ratheesh}
\affiliation{INAF -- IAPS, via Fosso del Cavaliere, 100, Rome, Italy I-00133}
\affiliation{Università di Roma “La Sapienza”, Dipartimento di Fisica, Piazzale Aldo Moro 2, Rome, Italy I-00185}
\affiliation{Università di Roma "Tor Vergata", Dipartimento di Fisica', Via della Ricerca Scientifica, 1, I-00133 Roma, Italy}

\author{Alda Rubini}
\affiliation{INAF -- IAPS, via Fosso del Cavaliere, 100, Rome, Italy I-00133}

\author{Francesco Santoli}
\affiliation{INAF -- IAPS, via Fosso del Cavaliere, 100, Rome, Italy I-00133}

\author{Paolo Sarra}
\affiliation{OHB Italia, Via Gallarate 150, I-20151 Milano, Italy}

\author{Emanuele Scalise}
\affiliation{INAF -- IAPS, via Fosso del Cavaliere, 100, Rome, Italy I-00133}

\author{Andrea Sciortino}
\affiliation{OHB Italia, Via Gallarate 150, I-20151 Milano, Italy}

\author{Carmelo Sgrò}
\affiliation{INFN-Pisa, Largo B. Pontecorvo 3, I-56127 Pisa, Italy}

\author{Gloria Spandre}
\affiliation{INFN-Pisa, Largo B. Pontecorvo 3, I-56127 Pisa, Italy}

\author{Marcello Tardiola}
\affiliation{OHB Italia, Via Gallarate 150, I-20151 Milano, Italy}

\author{Allyn F. Tennant}
\affiliation{NASA Marshall Space Flight Center, Huntsville, AL 35812, USA}

\author{Antonino Tobia}
\affiliation{INAF -- IAPS, via Fosso del Cavaliere, 100, Rome, Italy I-00133}

\author{Alessio Trois}
\affiliation{INAF-OAC, Via della Scienza 5, I-09047 Selargius (CA), Italy}

\author{Marco Vimercati}
\affiliation{OHB Italia, Via Gallarate 150, I-20151 Milano, Italy}

\author{Martin C. Weisskopf}
\affiliation{NASA Marshall Space Flight Center, Huntsville, AL 35812, USA}

\author{Davide Zanetti}
\affiliation{INFN-Pisa, Largo B. Pontecorvo 3, I-56127 Pisa, Italy}

\author{Francesco Zanetti}
\affiliation{OHB Italia, Via Gallarate 150, I-20151 Milano, Italy}

\begin{abstract}

IXPE (Imaging X-ray Polarimetry Explorer) is a NASA Small Explorer mission -- in partnership with the Italian Space Agency (ASI) -- dedicated to X-ray polarimetry in the 2--8 keV energy band. The IXPE telescope comprises three grazing incidence mirror modules coupled to three detector units hosting each one a Gas Pixel Detector (GPD), a gas detector that allows measuring the polarization degree by using the photoelectric effect. A wide and accurate ground calibration was carried out on the IXPE Detector Units (DUs) at INAF-IAPS, in Italy, where a dedicated facility was set-up at this aim. In this paper, we present the results obtained from this calibration campaign to study the IXPE focal plane detector response to polarized radiation. In particular, we report on the modulation factor, which is the main parameter to estimate the sensitivity of a polarimeter.

\end{abstract}

\keywords{X-rays --- polarimetry --- gas detectors}

\section{Introduction} \label{sec:intro}

The Imaging X-ray Polarimetry Explorer (IXPE) \citep{ixpe1,ixpe2,ixpe3,ixpe4,ixpe5,Soffitta21,weisskopf2021} will be the first X-ray astronomy mission fully dedicated to polarimetry; it will expand our knowledge on X-ray sources, adding polarization data to temporal, spectral, and imaging ones allowing to obtain scientifically relevant measurements from several sources (e.g., neutron stars, black-holes, AGNs, SNRs, ...). Despite the importance of polarimetric information, the only available statistically significant measurements of X-ray polarization were obtained for the Crab Nebula \citep{crab1,crab2} over 40 years ago by using the crystal polarimeters aboard the Orbiting Solar Observatory 8 (OSO-8) and recently by PolarLight \citep{feng} albeit with a much smaller significance\footnote{Recently also a low significance measurement of polarization of Sco-X1 has been performed by PolarLight \citep{ScoX1}.}. 

IXPE has been launched on 9$^{th}$ December 2021 into a near-equatorial circular orbit at about 600 km altitude. The IXPE Payload comprises three X-ray telescopes, each one with a X-ray optics and one DU separated by a shared optical bench (boom), deployed to match the telescopes' focal length. Each DU lid has mounted on a collimator, an ions-UV filter \citep{UV}. Each DU is equipped with a Filter and Calibration Wheel \citep{Ferrazzoli, ICE}, the Gas Pixel Detector (GPD) \citep{gpd1,gpd2,gpd3,Sgro} and the Back-End Electronics (BEE) \citep{Bee} which connects to the Detector Service Unit (DSU). The whole IXPE instrument has been built by INAF--IAPS and INFN and it is described in more detail in \cite{Soffitta21}.

\begin{figure*}[!h]
	\gridline{\fig{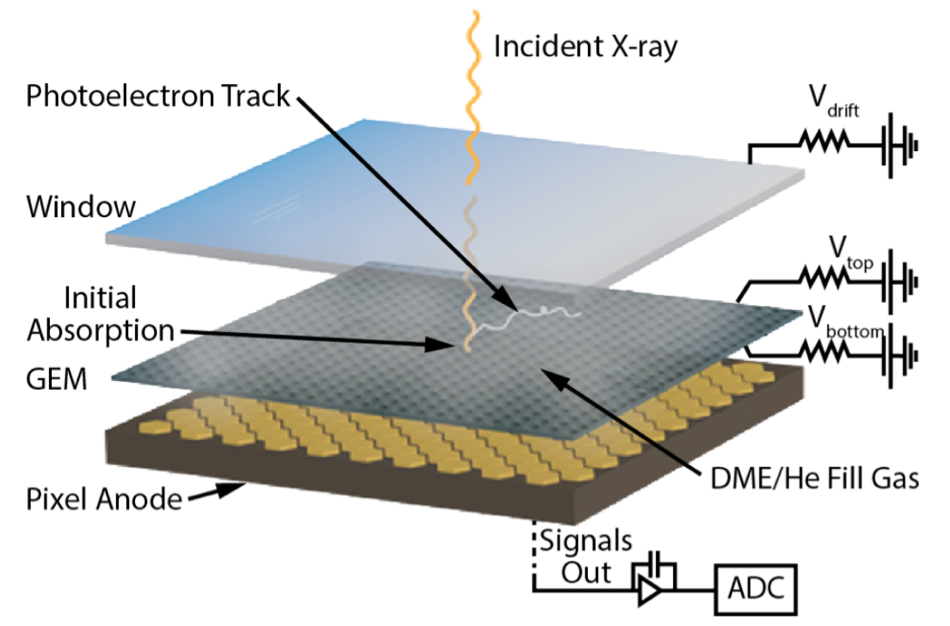}{0.35\textwidth}{(a)}
		\fig{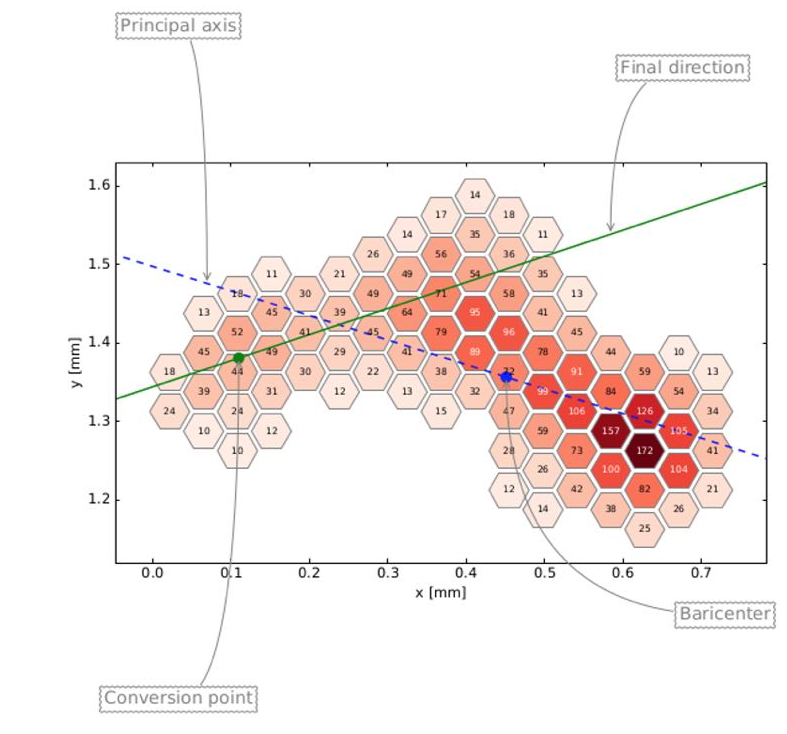}{0.35\textwidth}{(b)}
	}
	\caption{A schematic view of the Gas Pixel Detector (GPD) is shown on the (a) panel, (b) panel displays an ionization track resulting from absorption of a 5.9 keV X ray imaged onto the GPD's pixelated anode \citep{ixpe6}.
		\label{fig:gpd}}
\end{figure*}

\begin{figure}[!b]
	\centering
	\includegraphics[width=7cm]{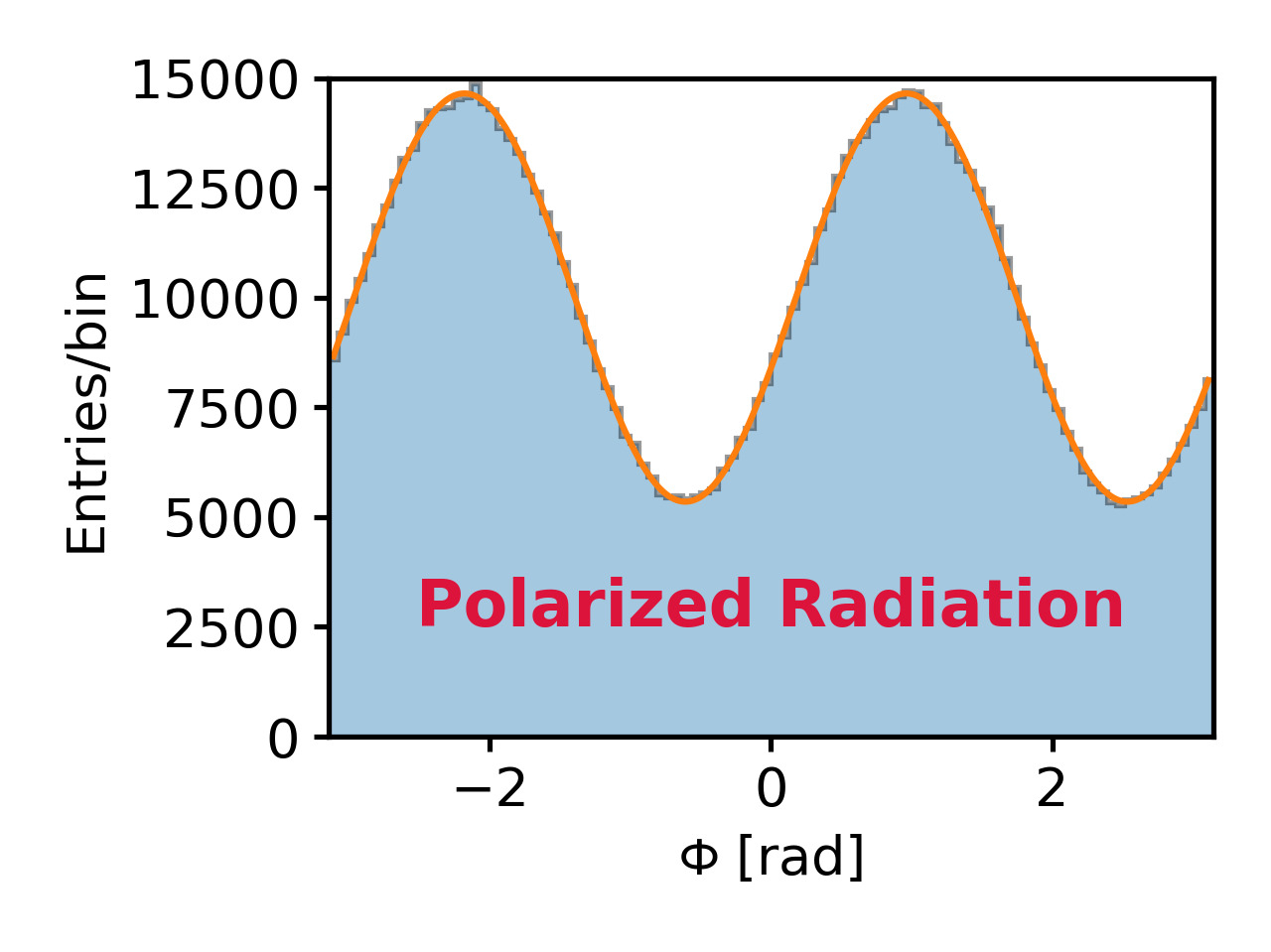}
	\includegraphics[width=7cm]{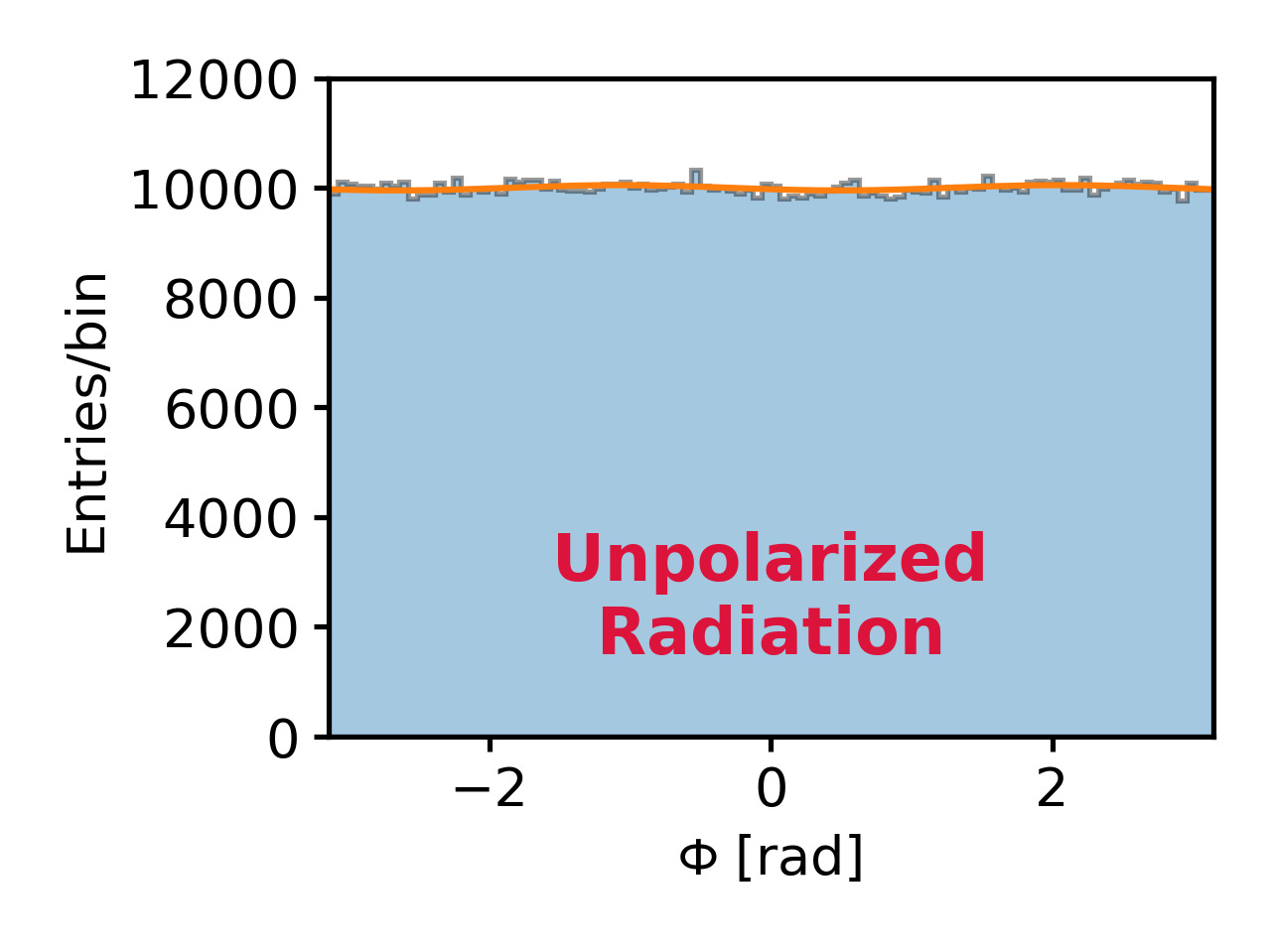}
	\caption{Distribution of photoelectrons directions (modulation curves) produced by polarized (left) and unpolarized (right) radiation at 5.89 keV \citep{Rankin2021}.
	}
	\label{fig:mcurves}
\end{figure}

The GPD was invented and developed by the IXPE Italian team, and it allows to obtain an image of the ionization track produced by the photoelectron resulting from absorption of an X-ray in the gas cell (see Figure \ref{fig:gpd}a).
The imaged ionization tracks (see the example in Figure \ref{fig:gpd}b) contains information on the photoelectron's energy and direction, which is correlated with the polarization orientation of the absorbed X-ray (in case of orbital $s$ electrons) as described by the differential cross section \citep{Heitler}:
\begin{equation}
\frac{d\sigma}{d\Omega}=r_0^2 \frac{Z^5}{137^4}\left( \frac{mc^2}{h\nu} \right)^{7/2} \frac{4\sqrt{2} \sin^2(\theta) \cos^2(\phi)}{\left[ 1- \beta \cos(\theta)\right]^4},
\label{eq:phe}
\end{equation}
where $r_0$ is the classical electron radius, Z the atomic number of the gas, $mc^2$ is the rest electron mass, $h\nu$ is the photon energy,  $\beta$ the fraction of electron velocity with respect to the speed of light, $\theta$ and $\phi$ are the polar and azimuthal angles respectively. Then in a photoelectric polarimeter, the X-rays interaction produces mainly photoelectrons with a $\cos^2 (\phi)$ angular distribution. As shown in Figure \ref{fig:mcurves}, photoelectrons angular distribution for a polarized (left) X-source show a  $\cos^2 (\phi)$ modulation that is not observable in the case of an unpolarized source (right).
The modulation amplitude, $M$, of the modulation curves is proportional to the polarization degree of the source, $P$. To obtain $P$ from a measured $M$ we need a parameter, named modulation factor, $\mu$:
\begin{equation}
P=\frac{M}{\mu}.
\end{equation}
The modulation factor is a property of the detector (the most important parameter for the calculation of the sensitivity for a polarimeter) and is given by the modulation amplitude measured in the presence of a 100\% polarized source.

Being IXPE a discovery mission, standard celestial sources are not available for performing in-flight calibration for polarimetry because the few available measurements of polarization obtained in the past \citep{crab1, crab2, feng, ScoX1} cannot be used as a flight calibrator, also because the source is known to vary. Because gas detectors based on GEM can have a time dependent response, IXPE is equipped with an on-board calibration system and a detailed ground calibration was mandatory. At this aim, a wide calibration campaign has been performed at INAF-IAPS in Rome. During such campaign, data were acquired 24 hours per day and 7 days per week: for each DU at least 40 days of calibrations were needed. Each of the four DUs have been calibrated separately (3 DU to be integrated on the payload plus a spare unit). Calibration results have also been validated by telescope calibrations, as cited in \cite{ixpe5,bongiorno}. 

In the following, chapters the set-up used during the calibration campaign is briefly presented and the used polarized sources are summarized, a complete description is given in \cite{Muleri21}. Below, the results obtained from different methods of analysis for the modulation factor are compared, then following the approach of \cite{DiMarco21}, the modulation factor has been widely characterized for all the IXPE DUs and results are compared with IXPE scientific requirements.

\section{Experimental set-up}

\begin{figure}[!b]
	\centering
	\includegraphics[height=6cm]{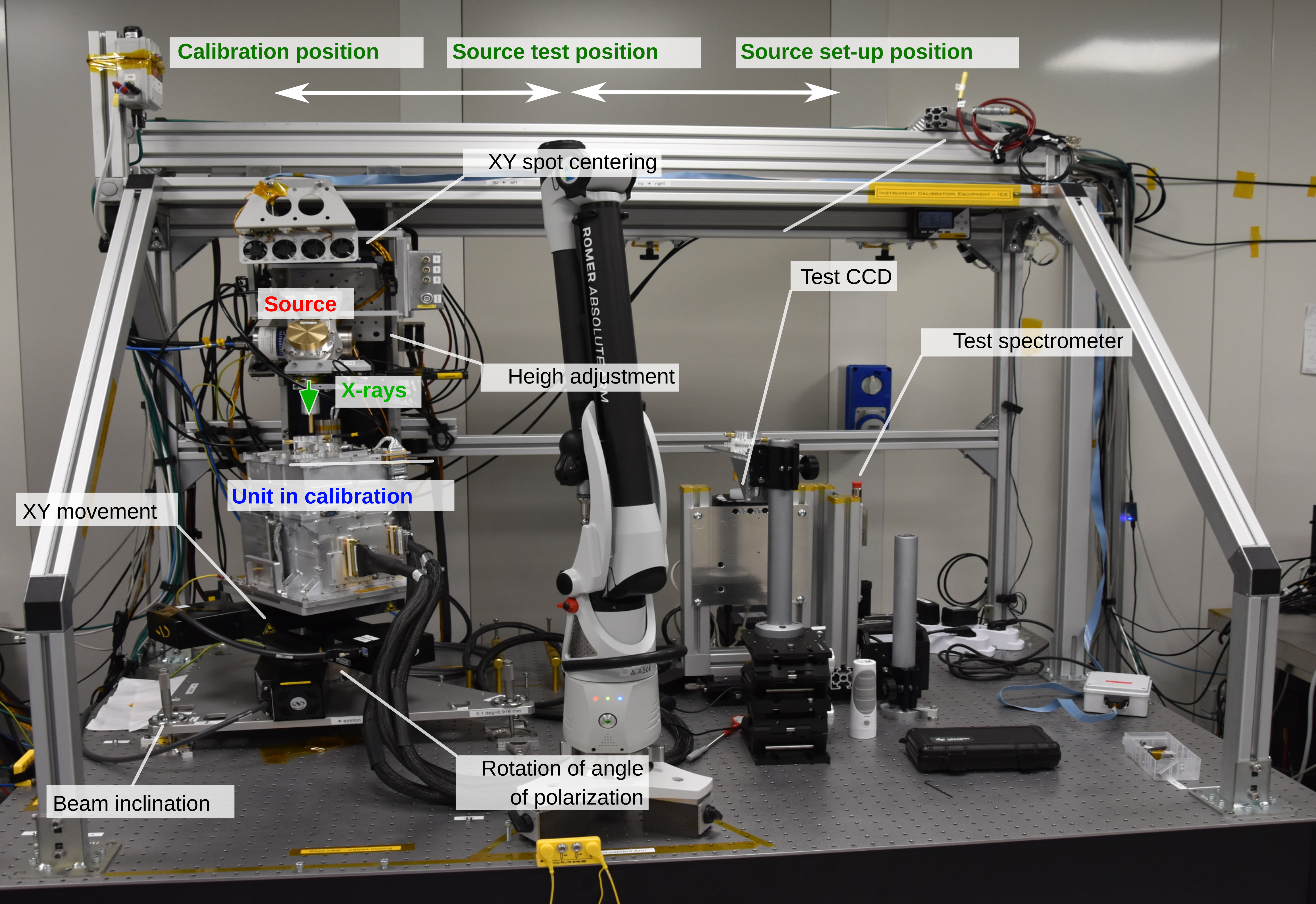}
	\includegraphics[height=6cm]{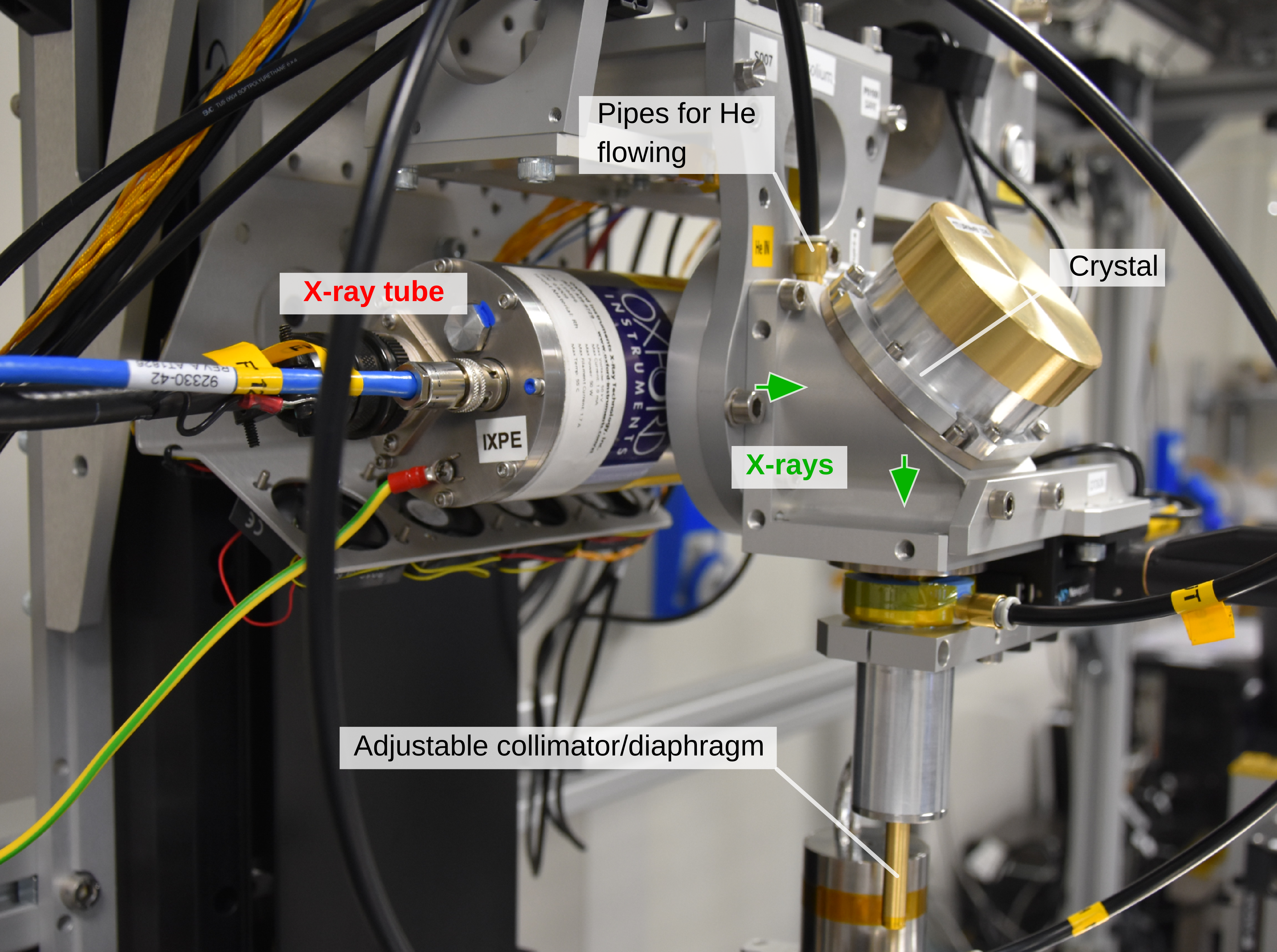}
	\caption{$Left$: Frontal view of the ICE during a measurement with a DU-FM. $Right$: view of one of the polarized sources set-up \cite{Muleri21}.
	}
	\label{fig:ice}
\end{figure}

An extensive X-ray calibration has been performed at INAF/IAPS for all the IXPE DUs; such calibrations have been carried out with a set-up named Instrument Calibration Equipment (ICE, see Figure \ref{fig:ice}) constructed at this aim in an ISO7 (10,000 class) clean room \citep{Muleri21}. 
IXPE four DU Flight Model (DU-FM), three to be installed on the payload and one spare unit, have been calibrated with the same procedures and the same experimental measurements have been performed. The DU-FMs are named with numbers from 1 to 4 (the number 1 is the spare unit) and they were calibrated in the order: DU-FM2, DU-FM3, DU-FM4, DU-FM1. The calibration equipment includes: 
\begin{itemize}
	\item X-ray sources used for calibration and tests (each source emits X-ray photons at known energy and with known polarization degree and angle);
	\item control system that allows knowning the direction of the beam, the direction of polarization for polarized sources, and its position with respect to the GPD inside the DU-FM (so it can be aligned and moved as necessary);
	\item the test detectors (SDD and X-ray CCD camera) which are used to characterize the beam before DU-FM calibration and as a reference for specific measurements;
	\item electrical and mechanical ground support equipment required to operate the DU-FMs.
\end{itemize}

On the ICE set-up, X-ray sources are mounted on a platform which can slide in three different positions: one for setting-up the source, one for testing it with test detectors and one to calibrate the DU-FM. Controlled independently stages allow to align the source to the DU-FM and to move it with respect to the source during calibration. These stages allow controlling the geometry of the set-up with a sensitivity at order of 10 $\mu$m. 

At INAF-IAPS both polarized and unpolarized sources are available \citep{Muleri21} and have been used during ground calibrations. Measurements with sources at several energies have been carried out to study the response to polarized radiation. The used polarized sources are based on commercial X-ray tubes (Oxford Series 5000 or Hamamatsu Head-on N7599 series) and Bragg diffraction at nearly 45$^\circ$ on a crystal, as listed in Table \ref{table:sources}.
\begin{table} [h]
	\begin{tabular}{l|l|l|l|l|l|l}
		Crystal & X-ray tube & Energy & 2d [$\AA$] & Diffraction & Rp/Rs & Polarization \\ 
		&            & [keV]  &            & angle [deg] &       &  [\%] \\ \hline
		PET (002) & Continuum & 2.01 & 8.742 & 45.00 & 0.0000 & $\simeq$100.0\% \\
		InSb (111) & Mo (L$\alpha$) & 2.29 & 7.481 & 46.28 & 0.0034 & 99.3\% \\
		Ge (111) & Rh (L$\alpha$) & 2.70 & 6.532 & 44.73 & 0.0024 & 99.5\% \\
		Si (111) & Ag (L$\alpha$) & 2.98 & 6.271 & 41.49 & 0.0252 & 95.1\% \\
		Al (111) & Ca (K$\alpha$) & 3.69 & 4.678 & 45.90 & 0.0031 & 99.4\% \\
		Si (220) & Ti (K$\alpha$) & 4.51 & 3.840 & 45.73 & 0.0023 & 99.5\% \\
		Si (400) & Fe (K$\alpha$) &	6.40 & 2.716 & 45.51 &   - -  & $\simeq$100.0\% \\ \hline
	\end{tabular}
	\caption{List of polarized sources set-ups used during DU-FM ground calibrations \cite{ICE,Muleri21}.}
	\label{table:sources}
\end{table}
A collimator is used with aim to obtain a beam with a well-defined and measurable direction and polarization angleand to select a portion of the diffracted and diverging beam as shown in Figure \ref{fig:ice}-$Right$. A nearly flat illumination over a region is obtained translating the DU continuously on a plane perpendicular to the source beam during measurement, the algorithm used is the same as the one proposed for dithering the IXPE satellite pointing; typically, the measurement with nearly 100\% polarized X-ray sources are repeated at 5 different polarization angles to verify that modulation factor is independent on the polarization angle as expected; moreover two of them are rotated 90$^\circ$ relative to each other to allow to estimate systematic effects. Analyzed data sets comprises:
\begin{itemize}
	\item Polarized Flat Field (PolFF) at seven energies (2.01, 2.29, 2.70, 2.98, 3.69, 4.51 and 6.40 keV). These sources are monochromatic for all practical purposes and produced by a spot which is moved to illuminate a central circular area with 7 mm radius. With Flat Field is intended a measurement covering the whole sensitive area of the detector. The rotation of the DU, named $\epsilon_1$, is the same at all energies and it corresponds to have an expected polarization angle value of $\simeq$55$^\circ$.
	\item Polarized Deep Flat Field (PolDFF) at seven energies, same sources used for PolFFs, where the spot is moved on a circular area at the center with 3.25 mm radius to collect more events per cm$^2$ in a shorter time. With Deep Flat Field is intended a measurement covering a smaller region on the  sensitive area of the detector, but with an high count rate. PolDFF are repeated at different polarization angles and the number of measurements depended on the source counting rate (sources at 2.01 and 3.69 keV having lower rate were carried out only to an angle in the DU-FM frame $\epsilon_2$=-35$^\circ$, the other ones having higher rate at four angles $\epsilon_2$=32.5$^\circ$, $\epsilon_3$=10$^\circ$, $\epsilon_4$=-12.5$^\circ$ and $\epsilon_5$=-35$^\circ$).
\end{itemize}

The analysis in the following are performed: (i) to verify polarimetric response uniformity of the detectors studying the FF data; (ii) after uniformity confirmation, to obtain the most sensitive estimation of the modulation factor, analysis are performed in a circular centered region with radius 3.0 mm, that is the largest region illuminated at every energy and angle during ground calibration with polarized sources.

\section{Analysis method}\label{modfact}

In literature, several methods to estimate the polarization degree and angle from data have been proposed. In the following, we recall three of them: (i) the classical one based on modulation amplitude estimation from modulation curves fit; (ii) Stokes parameters estimation from modulation curves fit \cite{strohmayer}; (iii) Stokes parameters estimation based on an unbinned event-by-event approach \cite{Kislat}. The three methods have been applied to the same data-set obtaining every time the same result, in the folowing they are shortly presented and applied, as an example, to data acquired with the DU-FM2 in the presence of a 100\% polarized source at 2.7 keV and 6.4 keV as representative cases at low and high energy in the IXPE band.

\subsection{``Classical'' approach}\label{classical}

From Equation \ref{eq:phe} and Figure \ref{fig:mcurves}--left, it is evident that the modulation curve, in presence of a polarized radiation, follows a $\cos^2 \phi$ distribution. The classical approach to obtain polarimetric information consists in fitting this distribution with the function:
\begin{equation}\label{eq:classical}
\mathcal{M}(\phi)=A+B \cos^2(\phi - \phi_0),
\end{equation}
from which the modulation is derived:
\begin{equation}
M=\frac{\mathcal{M}_{max} - \mathcal{M}_{min}}{\mathcal{M}_{max}+\mathcal{M}_{min}}=\frac{B}{2A+B}.
\end{equation}

In this approach, the polarization angle is given directly by the fit as $\phi_0$ and the polarization degree is given by:
\begin{equation}
P=\frac{1}{\mu}\frac{B}{2A+B}.
\end{equation}

\begin{figure}[!h]
	\centering
	\includegraphics[width=7cm]{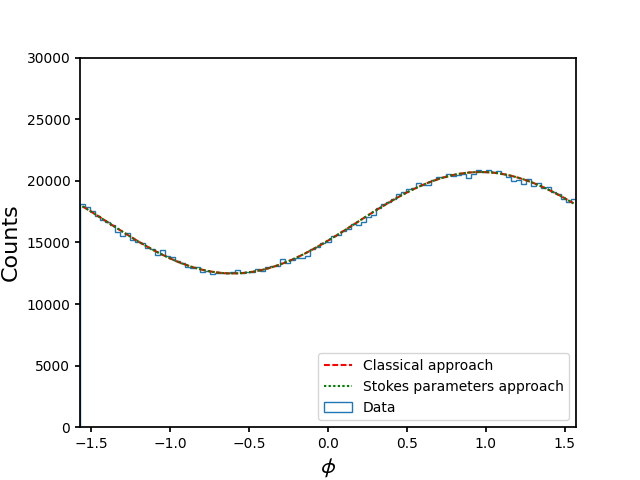}
	\includegraphics[width=7cm]{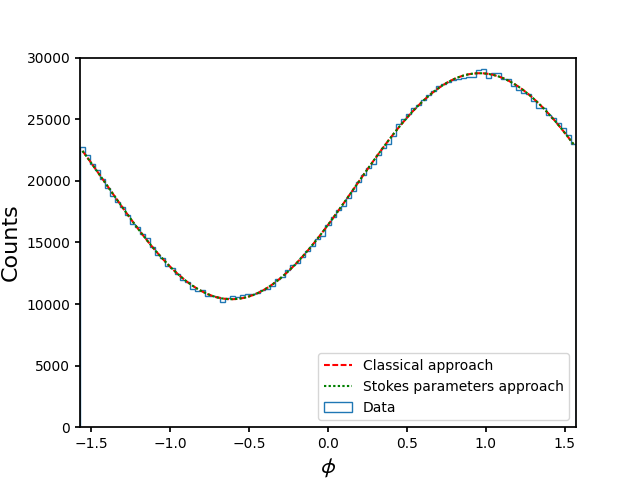}	
	\caption{Example of fit with the classical approach (see section \ref{classical}) and with the Stokes parameters approach (see section \ref{stokes}) for events at 2.7 keV ($Left$) and 6.4 keV ($Right$). The two best--fit curves are perfectly overimposed.
	}
	\label{fig:fit_comp}
\end{figure}

In Figure \ref{fig:fit_comp} the modulation curves at 2.7 keV ($Left$) and 6.4 keV ($Right$) are shown superimposed to the best--fit curves from this method (red-dashed lines). The obtained value for the modulation amplitude is reported in Table \ref{tab3}.
\begin{table}[!th]
	\begin{tabular}{c|c|c|c|c}
		Energy & Parameter & Classical Approach & Stokes parameters & Event-by-event \\ \hline
		2.7 keV & $\mu$ & $(24.80 \pm 0.14)$\% & $(24.796 \pm 0.084)$\% & $(24.81 \pm 0.11)$\% \\ \cline{2-5}
		& $\phi$ & -(34.67$\pm$0.17)$^\circ$ & -(34.67$\pm$0.10)$^\circ$ & -(34.69$\pm$0.13)$^\circ$ \\ \hline
		6.4 keV & $\mu$ & $(46.90 \pm 0.11)$\% &  $(46.896 \pm 0.067)$\% & $(46.92 \pm 0.10)$\% \\ \cline{2-5}
		& $\phi$ & -(34.990$\pm$0.074)$^\circ$ & -(34.990$\pm$0.045)$^\circ$ & -(34.989$\pm$0.062)$^\circ$ \\ \hline
	\end{tabular}\caption{\label{tab3}
		Modulation factor and polarization angle measured, for the same data at 2.7 keV and 6.4 keV, using the three different analysis methods.}
\end{table}

\subsection{Binned Stokes parameters approach}\label{stokes}

An alternative approach consists in using Stokes parameters as described in \cite{strohmayer}. The polarization degree as a function of the Stokes parameters is:
\begin{equation}
P=\frac{\sqrt{U^2+Q^2}}{\mu I},
\end{equation}
thus for a 100\% linearly polarized beam $I=\sqrt{U^2 + Q^2}$. It is convenient to use the normalized Stokes parameters, $q=Q/I$ and $u=U/I$, therefore the degree of polarization and the direction are:
\begin{eqnarray}\label{eq:stokes}
P & = & \frac{\sqrt{q^2 + u^2}}{\mu} \nonumber \\
\phi_0 & = & \frac{1}{2} \arctan \left( \frac{u}{q} \right).
\end{eqnarray}

In this approach, proposed in \cite{strohmayer}, modulation curves can be fitted with the function:
\begin{equation}\label{eq:stromayer}
\mathcal{M}(\phi) = I + Q \cos (2\phi) + U \sin (2\phi) ,
\end{equation}
allowing us to obtain the Stokes parameters directly from the fit. In Figure \ref{fig:fit_comp} the best--fit curves are shown in green-dotted lines, the modulation amplitude and phase are reported in Table \ref{tab3}.

\subsection{Unbinned Stokes parameters approach}\label{eventbyevent}

For each detected event, IXPE DUs provide the photoelectron emission direction $\phi_k$; this allows to estimate for each one of them the Stokes parameters following the approach proposed by \cite{Kislat}:
\begin{eqnarray}\label{eq:kislat}
i_k & = & 1 \nonumber \\
q_k & = & 2\cos(2\phi_k) \\
u_k & = & 2\sin(2\phi_k). \nonumber
\end{eqnarray}
From these unbinned Stokes parameters, it is possible to determine the ones for an observation of N events:
\begin{eqnarray}
I & = & \sum_N i_k = N \nonumber \\
Q & = & \sum_N q_k = Nq \\\nonumber 
U & = & \sum_N u_k = Nu
\end{eqnarray}
with associated uncertainties (in case of small polarization degree):
\begin{eqnarray}
\sigma_I & =      & \sqrt{N} \nonumber \\
\sigma_q & \simeq \sigma_u \simeq & \sqrt{\frac{2}{N-1}} \nonumber 
\end{eqnarray}

The polarization degree and angle can be derived from Equations \ref{eq:stokes}. In this approach there is not a fit, but only a numerical estimation that has been reported in Table \ref{tab3}.

\subsection{Adopted analysis approach}

In Table \ref{tab3} the modulation factor and the polarization angle from the three methods are compared, they give rise to compatible values for both modulation factor and polarization angle. Stokes parameters have the great advantage of being handled as fluxes with full poissonian statistics. As an example, this allows decoupling systematic effects in polarization measurements. Moreover an unbinned analysis allows to obtain estimations also from a low number of events and to apply a weighted analysis as described in \cite{DiMarco21}. Moreover, a polarimeter can suffer from a systematic named spurious modulation, i.e. the presence of a small modulation due to the instrument, also present when unpolarized sources are observed. This effect can be calibrated and subtracted to correct the polarization degree of the source following a correction algorithm as the one described in \cite{Rankin2021}. 

\subsubsection{Spurious modulation subtraction}

Two existing approaches based on the use of Stokes parameters can be applied to disentangle the spurious component from the source. The first approach decouples the two measurements acquired with the detector rotated of an angle $\epsilon_2 - \epsilon_1 = \Delta \epsilon = 90 ^\circ$. Calling $(q_{spurious}, u_{spurious})$ and $(q_{source}, u_{source})$ the Stokes parameters for the instrument response due to spurious effects and the source, respectively, Stokes parameter q (or u) measured at the first and second angles read:
\begin{eqnarray}
q_{1} & = & q_{spurious} (\epsilon_1) + q_{source} (\epsilon_1) \\
q_2 & = & q_{spurious} (\epsilon_2) + q_{source} (\epsilon_2) = q_{spurious} (\epsilon_1 + 90 ^\circ) + q_{source} (\epsilon_1 + 90^\circ) = \nonumber \\ 
& = & q_{spurious} (\epsilon_1) - q_{source} (\epsilon_1) \\ \nonumber
\end{eqnarray}
and then
\begin{equation}
\left\{ \begin{array}{l}
q_{spurious}=\frac{q_1+q_2}{2} \\
q_{source}=\frac{q_1-q_2}{2} \\
\end{array} \right.
\end{equation}
A second approach is based on calibration maps, obtained from ground measurements. For each observed event in a given GPD pixel with coordinates ($x_k, y_k$) and energy $E_k$, it is possible to subtract the spurious modulation value as explained in \cite{Rankin2021}:
\begin{eqnarray}
i_{cal} & = & i_k \nonumber \\
q_{cal} & = & q_k - q_{sm}(x_k,y_k,E_k)  \\ \nonumber
u_{cal} & = & u_k - u_{sm}(x_k,y_k,E_k).  
\end{eqnarray}

\subsubsection{Modulation factor estimation}
In the past, GPD data analyses have been carried out by applying a selection of the events named ``Standard cuts'' \citep{Muleri16,DiMarco21}, where about 20\% of the events are removed. The selection is applied in two steps: i) remove events with energy values outside $\pm 3 \sigma$ from the peak centroid; ii) remaining tracks are ordered for ``elongation'' (that is the ratio between the track length and track width, for more details see \cite{DiMarco21}) the ones with lower-elongation are removed up to reach a threshold that allows to removing 20\% of the initial events, including the ones removed at the step (i). This approach cannot be applied to astronomical sources which are not-monochromatic, thus a new approach to obtain an optimal sensitivity has been proposed in \cite{DiMarco21} consisting in applying for each event an optimal weight, $w_i$, to the Stokes parameters: 
\begin{eqnarray}
I&=&\sum_i w_i \nonumber \\
Q&=&\sum_i 2w_i \sin (2\phi_i) \\
U&=&\sum_i 2w_i \cos (2\phi_i) \nonumber 
\end{eqnarray}
where $w_i = \alpha^{0.75}$ with $\alpha$ given by the track ellipticity \cite{DiMarco21}). In the following, the weighted approach is applied after selection of events in a circular centered region with radius 3.0 mm.

\section{Calibration of response to polarization}

In this section, we present the analysis of the response to polarized sources for the IXPE DUs during ground calibrations.

\subsection{Uniformity of polarimetric response}

During ground calibrations, the X-ray sources are centered with the GPD. As a matter of fact, after the integration and in orbit deployment of mirrors, the focalized source beam could impinge on a different position, so that the observatory can dithers on a different portion of the GPD surface with respect to the one used for calibration. In this section the spatial uniformity of the polarimetric response is investigated. These studies will allow to asses that the results obtained in the central DFF region are valid for the whole GPD surface.

\begin{figure}[!h]
	\centering
	\includegraphics[width=8cm]{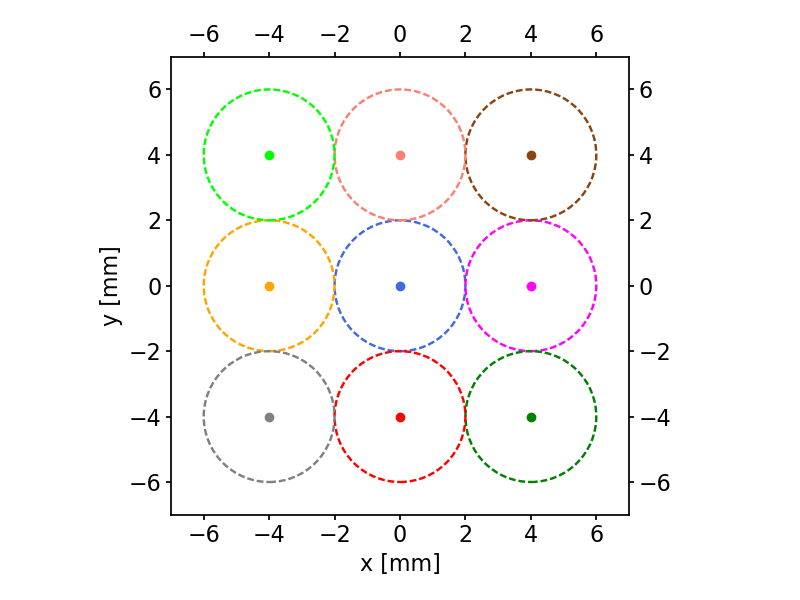}
	\caption{Independent regions in which the modulation factor is evaluated.}
	\label{fig:circle_map}
\end{figure}

To verify the spatial uniformity of the response to polarized radiation of the DUs, we performed an analysis on the FF data-sets at different energies, selecting 9 smaller, independent, regions with radius 2 mm (see Figure \ref{fig:circle_map}).
%

\begin{figure}[!h]
	\centering
	\includegraphics[width=7cm]{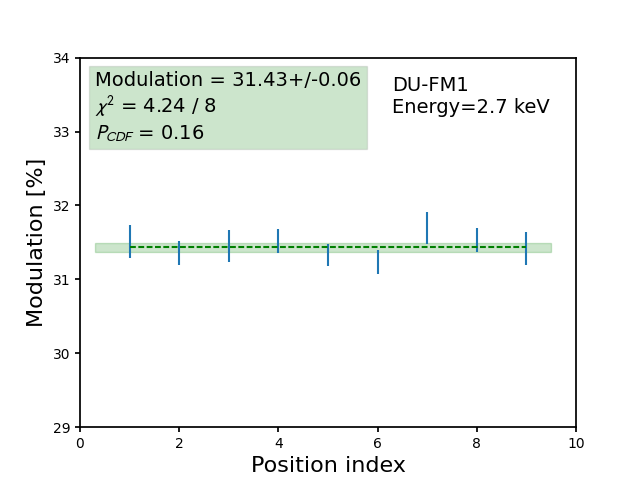}
	\includegraphics[width=7cm]{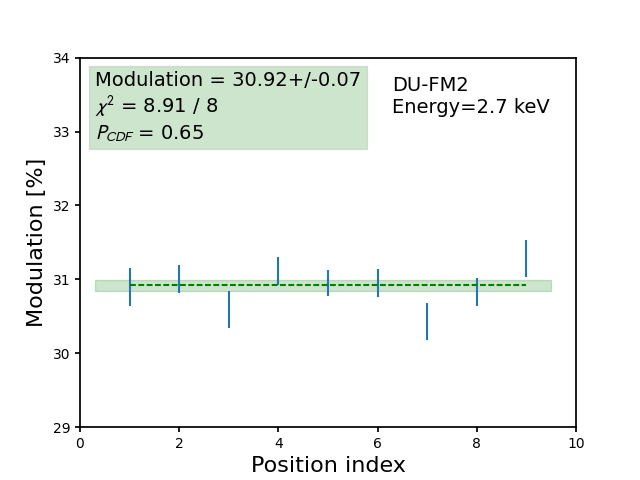}
	\includegraphics[width=7cm]{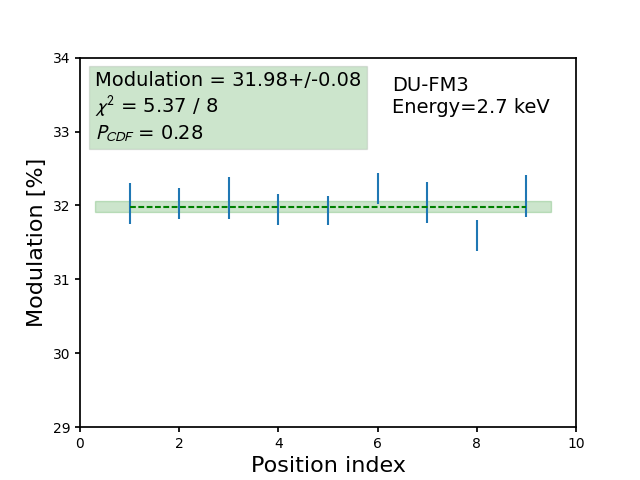}
	\includegraphics[width=7cm]{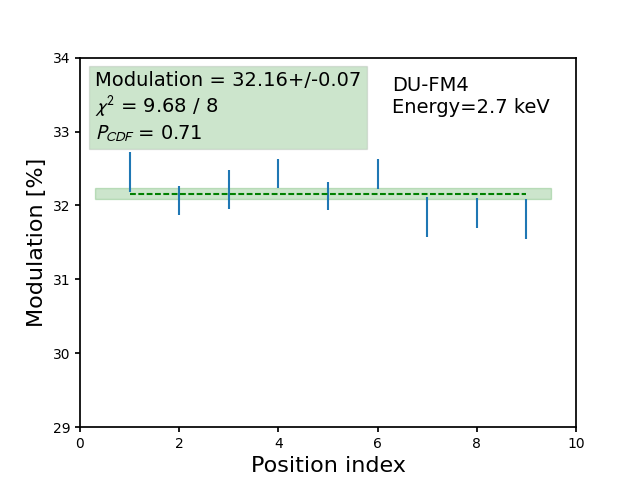}
	\caption{Modulation factor measured at 2.98 keV in 9 different spots of radius 2 mm with the DU-FM1 (top left), DU-FM2 (top right), DU-FM3 (bottom left) and DU-FM4 (bottom right).
	}
	\label{fig:uniformity3p69}
\end{figure}

\begin{figure}[!h]
	\centering
	\includegraphics[width=8cm]{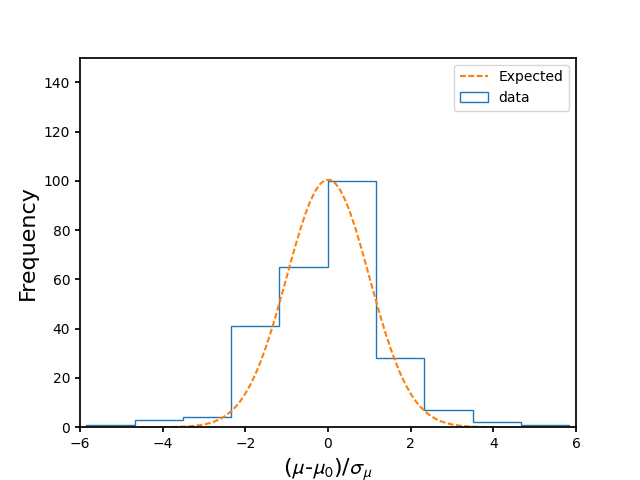}
	\caption{Distribution of the number of standard deviation, $\sigma_{\mu_i}$, between the best--fit mean value on all the 9 regions of the data-set ($\mu_0$) and the single measured modulation factor ($\mu_i$) for every energy and DU: $(\mu_i - \mu_0)/\sigma_{\mu_i}$. In $dashed$-orange the expected normal distribution is shown.
	}
	\label{fig:uniformity}
\end{figure}

In these regions the modulation factor has been evaluated and results allowed us to verify that the response of each DU is uniform on the whole sensitive area for each energy. For example in Figure \ref{fig:uniformity3p69} the measured modulation factors in the 9 regions for all the DUs are shown for the measurements at 2.98 keV and are compatible with a constant function.
%
For each position, we can estimate $(\mu_i - \mu_0)/\sigma_{\mu_i}$, that is the number of standard deviations ($\sigma_{\mu_i}$) between the measured modulation factor $\mu_i$ and the mean value ($\mu_0$) obtained fitting each data-set. The distribution of these values considering all the DUs and energies is shown in Figure \ref{fig:uniformity} and it is possible to observe that the distribution is normally distributed around zero.
%


\subsection{Dependance of modulation factor on the polarization angle} \label{mu}

\begin{figure}[!t]
	\centering
	\includegraphics[width=0.9\textwidth]{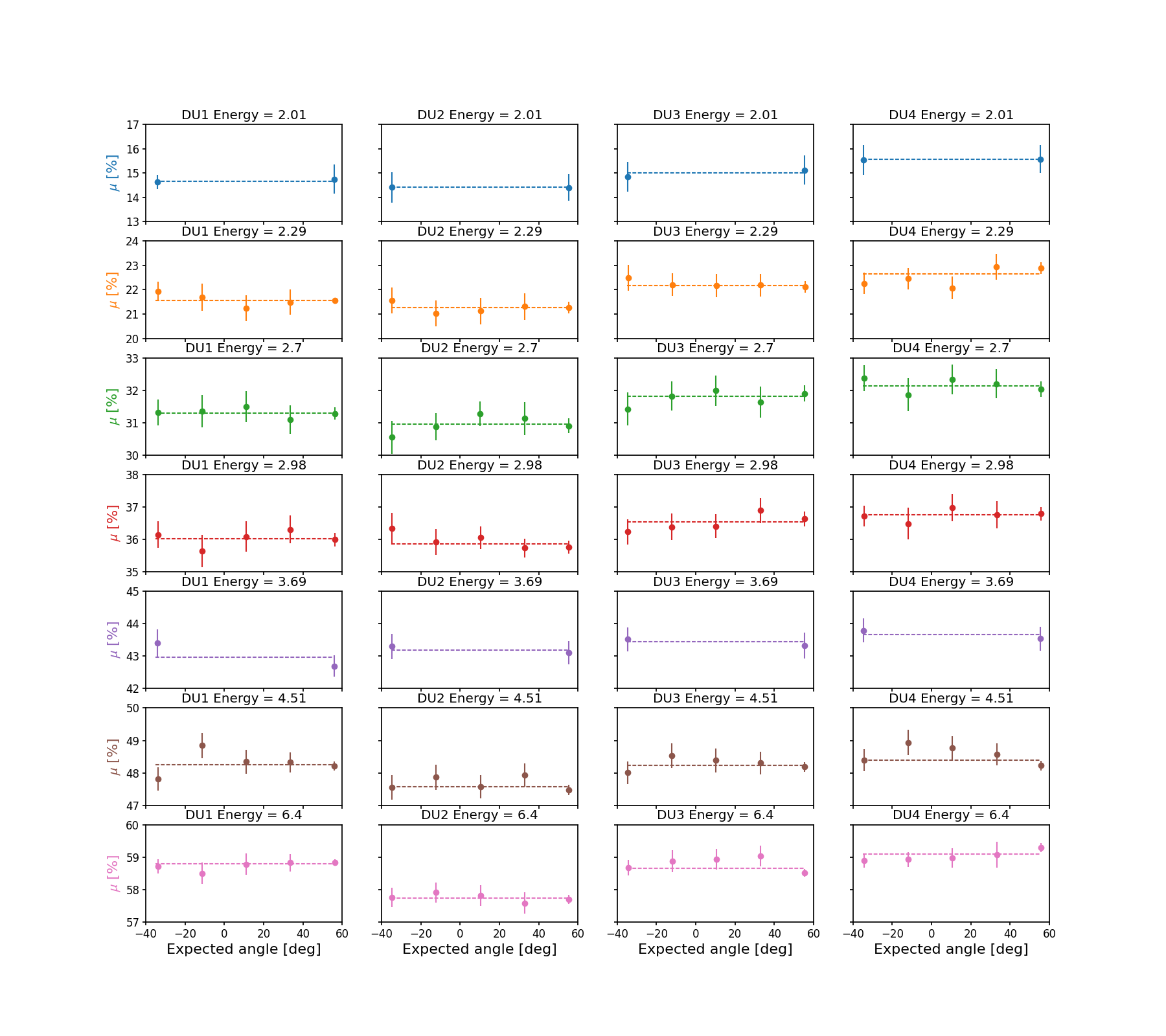}
	\caption{Modulation factor as a function of the expected polarization angle for the four DUs. Dashed lines represent the modulation factor mean values.}
	\label{fig11}
\end{figure}

On the IXPE focal plane, the three DUs will be clocked at 120$^\circ$ one with respect to the other, which means that the DUs will measure three values of the polarization angle of an astrophysical source shifted by 120$^\circ$ with respect to the detector coordinates. Such configuration allows for checking the genuine polarization from celestial sources. If the spurious modulation is properly removed, we expect that the modulation factor is independent of the polarization angle. Indeed the possible presence of an additional $\cos^2\phi$ spurious modulation contribution could modify the amplitude and phase of the modulation curve from a polarized source, hence the measurement of the modulation factor. In this section, we study the modulation factor as a function of the polarization angles. Results are shown in Figure \ref{fig11} for the four DUs at all the energies.

From this analysis results that the modulation factor does not depend on the polarization angle in the GPD frame once the spurious modulation is properly subtracted.

\vfill
\subsection{Absolute knowledge of polarization angle}\label{angle}

It is important for IXPE to determine precisely the polarization angle of a celestial source. This is because the polarization angle is correlated to the intrinsic geometry of the system under study and to the configuration of magnetic fields, such as in extended sources like SuperNova Remnants and Pulsar Wind Nebulae. A good knowledge of the position angle, also, it allows for comparing multi-wavelengths observations and detect secular variations. We recall that at level of the observatory the knowledge of the position angle is 1 degree \cite{Soffitta21}. Therefore, it is necessary to determine the polarization angle measurement accuracy with respect to the detector mechanical coordinates. For this reason, we used an Absolute Rohmer Arm with an accuracy of $\simeq 10 \mu$m, to map the geometry of the set-up and to correlate the measured polarization angle with the one expected from this mechanical set-up.
\begin{figure}[!t]
	\centering
	\includegraphics[width=0.9\textwidth]{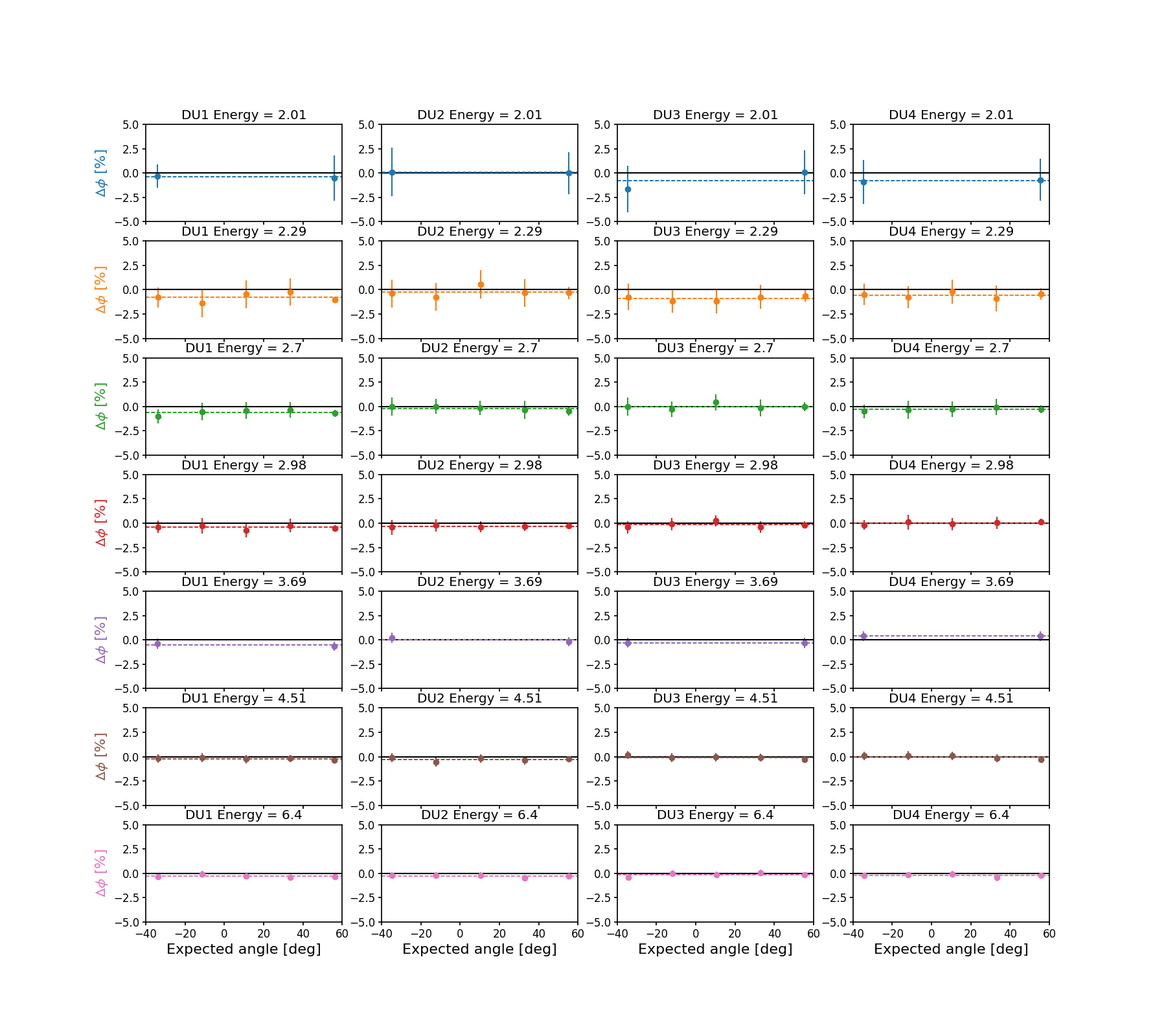}
	\vspace{-1cm}
	\caption{Deviation from the expected phase for polarization angle for each DU and energy. Data points are fitted with a constant line (dashed line).
	}
	\label{fig10}
\end{figure}
In this analysis, we estimated the polarization angle by the data, then its expected value is subtracted. Results from this analysis are shown in Figure \ref{fig10} for all the DUs and energies. In each box, points in color are the measured values fitted with a constant function (dashed line), to verify the ideal case (null-hypothesis).

For all the energies and DUs from the fit with a constant value an uncertainty on the polarization angle estimate has been determined. In particular at 6.4 keV the scientific requirement for IXPE, that is 0.4$^\circ$, is reached for all the DUs (see Table \ref{tab:syst}). These values state the level of accuracy that IXPE can reach for the polarization angle at each energy in each DU.
\begin{table}[!hb]
	\begin{tabular}{c|c|c|c|c}
\hline\hline
		Energy [keV] & DU-FM1 	  & DU-FM2 	& DU-FM3 	& DU-FM4 \\
		& $\Delta\phi$ [deg] & $\Delta\phi$ [deg] & $\Delta\phi$ [deg] & $\Delta\phi$ [deg] \\ \hline
		2.01 & -0.423 $\pm$ 0.073 & 0.047 $\pm$ 0.032 & -0.79 $\pm$ 0.61 & -0.802 $\pm$ 0.074 \\
		2.29 & -0.78 $\pm$ 0.18 & -0.24 $\pm$ 0.20 & -0.90 $\pm$ 0.11 & -0.56 $\pm$ 0.11 \\
		2.70 & -0.60 $\pm$ 0.10 & -0.201 $\pm$ 0.087 & -0.03 $\pm$ 0.11 & -0.293 $\pm$ 0.066 \\
		2.98 & -0.427 $\pm$ 0.079 & -0.320 $\pm$ 0.032 & -0.16 $\pm$ 0.11 & 0.006 $\pm$ 0.053 \\
		3.69 & -0.518 $\pm$ 0.092 & 0.03 $\pm$ 0.13 & -0.3053 $\pm$ 0.0051 & 0.3860 $\pm$ 0.0010\\
		4.51 & -0.211 $\pm$ 0.041 & -0.292 $\pm$ 0.078 & -0.074 $\pm$ 0.067 & -0.032 $\pm$ 0.066 \\
		6.40 & -0.288 $\pm$ 0.046 & -0.269 $\pm$ 0.049 & -0.129 $\pm$ 0.070 & -0.209 $\pm$ 0.055 \\ \hline
	\end{tabular}
\vspace{0.3cm}
	\caption{Systematic difference between the measured polarization angle and the expected one from the geometry of the experimental set-up for each IXPE DU-FM at every calibration energy. The values at 6.4 keV have to be compared with the scientific requirement of IXPE that is 0.4 deg. These values and errors are obtained from the linear fits of Figure \ref{fig10}.}
	\label{tab:syst}
\end{table}

\subsection{Modulation factor and angle systematic as a function of energy}

\begin{figure}[!h]
	\centering
	\includegraphics[width=10cm]{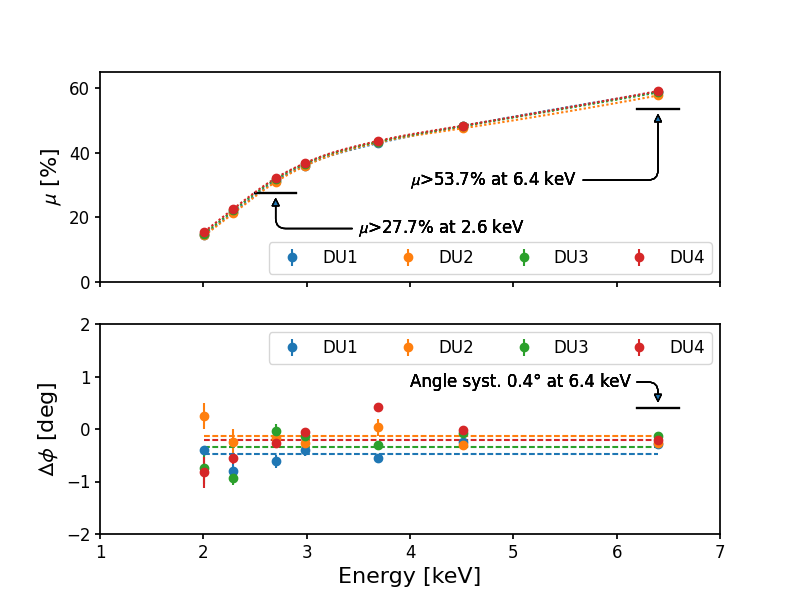}
	\caption{Modulation factor ($top$) and polarization angle systematic ($bottom$) as a function of energy. IXPE scientific requirements are reported in black. Mean values for the polarization angle systematic for each DU is reported as a dashed line.
	}
	\label{fig:mf_energy}
\end{figure}

The mean modulation factor estimated in Section \ref{mu} is plotted as a function of the energy for all the IXPE DUs in Figure \ref{fig:mf_energy}--$top$, where data points are fitted with a quadratic spline. In Figure \ref{fig:mf_energy}--$bottom$, the polarization angle uncertainty estimated in \ref{angle} is also reported as a function of the energy, the estimated values are reported in Table \ref{tab:syst}. 

Scientific requirements are reported on the plot of Figure \ref{fig:mf_energy} and are satisfied for all DUs.
The modulation factor values have been evaluated also using the unweighted approach of \cite{Kislat} and the Standard Cuts analysis, as shown in Table \ref{table:mu_comp}.

The modulation factor values and angle systematics for each DU are within the scientific requirement. The former is slightly different for the different DUs due to the small differences in the detectors \citep{Sgro} which are accounted for in the generation of the response matrices.

\begin{table}
\begin{tabular}{c|c|c|c|c}
\multicolumn{5}{c}{Weighted analysis}\\\hline\hline
Energy [keV] & DU-FM1 	  & DU-FM2 	& DU-FM3 	& DU-FM4 \\
& $\mu$ [\%]& $\mu$ [\%]	& $\mu$ [\%]	& $\mu$ [\%] \\ \hline
2.01 & 14.653 $\pm$ 0.048 & 14.4060 $\pm$ 0.0057 & 14.99 $\pm$ 0.14 & 15.557 $\pm$ 0.019 \\
2.29 & 21.567 $\pm$ 0.059 & 21.262 $\pm$ 0.067 & 22.182 $\pm$ 0.054 & 22.64 $\pm$ 0.16 \\
2.70 & 31.291 $\pm$ 0.045 & 30.949 $\pm$ 0.098 & 31.810 $\pm$ 0.086 & 32.128 $\pm$ 0.084 \\
2.98 & 36.028 $\pm$ 0.082 & 35.858 $\pm$ 0.088 & 36.55 $\pm$ 0.10 & 36.769 $\pm$ 0.057 \\
3.69 & 42.97 $\pm$ 0.35 & 43.19 $\pm$ 0.10 & 43.437 $\pm$ 0.093 & 43.66 $\pm$ 0.13 \\
4.51 & 48.26 $\pm$ 0.11 & 47.581 $\pm$ 0.085 & 48.229 $\pm$ 0.067 & 48.40 $\pm$ 0.12 \\
6.40 & 58.796 $\pm$ 0.042 & 57.729 $\pm$ 0.042 & 58.660 $\pm$ 0.094 & 59.103 $\pm$ 0.089 \\
\multicolumn{5}{c}{Unweighted analysis}\\\hline\hline
Energy [keV] & DU-FM1 	  & DU-FM2 	& DU-FM3 	& DU-FM4 \\
& $\mu$ [\%]& $\mu$ [\%]	& $\mu$ [\%]	& $\mu$ [\%] \\ \hline
2.01 & 12.644 $\pm$ 0.029 & 12.406 $\pm$ 0.092 & 12.827 $\pm$ 0.065 & 13.405 $\pm$ 0.040 \\
2.29 & 18.054 $\pm$ 0.044 & 17.803 $\pm$ 0.055 & 18.563 $\pm$ 0.088 & 18.960 $\pm$ 0.092 \\
2.70 & 25.817 $\pm$ 0.032 & 25.547 $\pm$ 0.066 & 26.169 $\pm$ 0.071 & 26.60 $\pm$ 0.10 \\
2.98 & 30.049 $\pm$ 0.064 & 29.877 $\pm$ 0.094 & 30.422 $\pm$ 0.076 & 30.771 $\pm$ 0.038 \\
3.69 & 36.99 $\pm$ 0.30 & 37.183 $\pm$ 0.096 & 37.37 $\pm$ 0.12 & 37.73 $\pm$ 0.18 \\
4.51 & 42.787 $\pm$ 0.097 & 42.165 $\pm$ 0.060 & 42.770 $\pm$ 0.054 & 42.98 $\pm$ 0.13 \\
6.40 & 54.614 $\pm$ 0.061 & 53.418 $\pm$ 0.060 & 54.46 $\pm$ 0.11  & 54.889 $\pm$ 0.081 \\
\multicolumn{5}{c}{Standard cuts analysis}\\\hline\hline
Energy [keV] & DU-FM1 	  & DU-FM2 	& DU-FM3 	& DU-FM4 \\
2.01 & 12.94 $\pm$ 0.12 & 12.68 $\pm$ 0.13 & 12.901 $\pm$ 0.098 & 13.653 $\pm$ 0.014 \\
2.29 & 20.582 $\pm$ 0.062 & 20.318 $\pm$ 0.072 & 21.221 $\pm$ 0.075 & 21.62 $\pm$ 0.18 \\
2.70 & 29.447 $\pm$ 0.036 & 29.156 $\pm$ 0.064 & 29.916 $\pm$ 0.088 & 30.348 $\pm$ 0.098 \\
2.98 & 34.099 $\pm$ 0.081 & 33.905 $\pm$ 0.090 & 34.466 $\pm$ 0.087 & 34.834 $\pm$ 0.065 \\
3.69 & 40.97 $\pm$ 0.39 & 41.23 $\pm$ 0.13 & 41.31 $\pm$ 0.18 & 41.60 $\pm$ 0.14 \\
4.51 & 46.38 $\pm$ 0.12 & 45.763 $\pm$ 0.078 & 46.236 $\pm$ 0.046 & 46.47 $\pm$ 0.13 \\
6.40 & 56.624 $\pm$ 0.044 & 55.987 $\pm$ 0.068 & 56.392 $\pm$ 0.062 & 56.89 $\pm$ 0.14 \\ \hline
\end{tabular}
\caption{Modulation factor evaluated with the weighted analysis of \cite{DiMarco21}, the unweighted one of \cite{Kislat} and the Standard Cuts. The best value is obtained using the weighted analysis.}
\label{table:mu_comp}
\end{table}

\subsection{Polarization response for off-axis beams}

\begin{figure}[!h]
	\centering
	\includegraphics[width=0.8\textwidth]{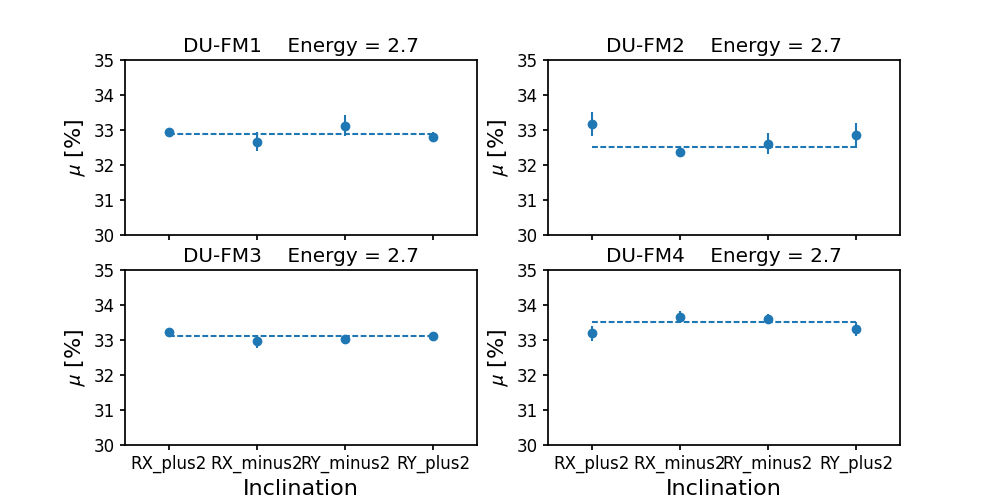}
	\caption{Modulation factor for the inclined measurements at 2.7 keV. They are compatible with a constant as expected.
	}
	\label{fig14}
\end{figure}

Photons focused by a mirror are incident on the detector at an angle of the order of a few degrees. In the case of the GPD, this causes a systematic effect as expected and computed in \cite{Muleri2014}. Albeit the effect is anticipated to be small, we verified such an expectation by measuring the modulation factor at 2.7 keV for a beam which is inclined of $\pm$2 degrees with respect to both axes normal the GPD window. This is the typical incidence angle from a focusing X-ray beam. Measurement is repeated at four azimuthal angles, to simulate the focusing from an IXPE mirror. The modulation factors for the four angles are shown in Figure \ref{fig14}. As expected, the modulation factors don't change with the azimuthal angle and the overall inclined modulation factor is compatible with the non-inclined one.

The modulation factors are constant within the uncertainties.

\section{Conclusions}
Being IXPE a discovery mission, known calibrated celestial sources are not available for performing in-flight calibration for polarimetry. In fact, the measurement performed 45 years ago on the Crab nebula \citep{crab1, crab2} and the recent ones by PolarLight \citep{feng,ScoX1} cannot be used as a flight calibrator. Moreover, these missions were not imaging, for these reasons a detailed ground calibration was mandatory. 

At this aim, a wide calibration campaign has been performed at INAF-IAPS. During such campaign data were acquired 24 hours per day and 7 days per week, for each DU at least 40 days of calibrations were needed. At the end of these calibrations, the results of this article were obtained for the IXPE polarimetric response, in particular:
\begin{itemize}
	\item  polarimetric response is uniform on the detector surface, once that spurious modulation has been subtracted;
	\item scientific requirements on the modulation factor at 2.6 keV and 6.4 keV are satisfied;
	\item once that the response to unpolarized radiation is calibrated, the modulation factor is constant with the polarization angle;
	\item systematic error on polarization angle at 6.4 keV does not exceeds 0.4 degrees;
	\item modulation factor does not depend on the X-ray beam inclination, thus it will be not affected by effects due to inclined penetration from the mirrors.
\end{itemize}

As a matter of fact, IXPE is equipped with both polarized and unpolarized sources to be used for checking the stability of the instrument's response during operation \citep{Ferrazzoli}. The response to polarized sources has been measured estimated with different analysis approaches that have been reported in Table \ref{table:mu_comp}. It is possible to observe that the weighted analysis of \cite{DiMarco21} allows to obtain the highest value of modulation factor. These results are used as a reference to obtain the IXPE CALDB distributed on HEASARC with the detectors response functions.

\section{Acknowledgments}
The Italian contribution to the IXPE mission is supported by the Italian Space Agency (ASI) through the contract ASI-OHBI-2017-12-I.0, the agreements ASI-INAF-2017-12-H0 and ASI-INFN-2017.13-H0, and its Space Science Data Center (SSDC), and by the Istituto Nazionale di Astrofisica (INAF) and the Istituto Nazionale di Fisica Nucleare (INFN) in Italy.

IXPE is a NASA Astrophysics Small Explorers (SMEX) mission, managed by MSFC and overseen by the Explorers Program Office at GSFC.

\bibliography{sample63}{}

\begin{thebibliography}{}
\expandafter\ifx\csname natexlab\endcsname\relax\def\natexlab#1{#1}\fi
\providecommand{\url}[1]{\href{#1}{#1}}
\providecommand{\dodoi}[1]{doi:~\href{http://doi.org/#1}{\nolinkurl{#1}}}
\providecommand{\doeprint}[1]{\href{http://ascl.net/#1}{\nolinkurl{http://ascl.net/#1}}}
\providecommand{\doarXiv}[1]{\href{https://arxiv.org/abs/#1}{\nolinkurl{https://arxiv.org/abs/#1}}}

\bibitem[{Baldini {et~al.}(2021)Baldini, Barbanera, Bellazzini, Bonino,
  Borotto, Brez, Caporale, Cardelli, Castellano, Ceccanti, Citraro, {Di Lalla},
  Latronico, Lucchesi, Magazzù, Magazzù, Maldera, Manfreda, Marengo,
  Marrocchesi, Mereu, Minuti, Mosti, Nasimi, Nuti, Oppedisano, Orsini,
  Pesce-Rollins, Pinchera, Profeti, Sgrò, Spandre, Tardiola, Zanetti, Amici,
  Andersson, Attinà, Bachetti, Baumgartner, Brienza, Carpentiero, Castronuovo,
  Cavalli, Cavazzuti, Centrone, Costa, D’Alba, D’Amico, {Del Monte}, {Di
  Cosimo}, {Di Marco}, {Di Persio}, Donnarumma, Evangelista, Fabiani,
  Ferrazzoli, Kitaguchi, {La Monaca}, Lefevre, Loffredo, Lorenzi, Mangraviti,
  Matt, Meilahti, Morbidini, Muleri, Nakano, Negri, Nenonen, O’Dell, Perri,
  Piazzolla, Pieraccini, Pilia, Puccetti, Ramsey, Rankin, Ratheesh, Rubini,
  Santoli, Sarra, Scalise, Sciortino, Soffitta, Tamagawa, Tennant, Tobia,
  Trois, Uchiyama, Vimercati, Weisskopf, Xie, Zanetti, \& Zhou}]{Sgro}
Baldini, L., Barbanera, M., Bellazzini, R., {et~al.} 2021, Astroparticle
  Physics, 133, 102628,
  \dodoi{https://doi.org/10.1016/j.astropartphys.2021.102628}

\bibitem[{Barbanera {et~al.}(2021)Barbanera, Citraro, Magazzù, Manfreda,
  Minuti, Nasimi, \& Sgrò}]{Bee}
Barbanera, M., Citraro, S., Magazzù, C., {et~al.} 2021, IEEE Transactions on
  Nuclear Science, 68, 1144, \dodoi{10.1109/TNS.2021.3073662}

\bibitem[{Bellazzini {et~al.}(2006)Bellazzini, Spandre, Minuti, Baldini, Brez,
  Cavalca, Latronico, Omodei, Massai, Sgro’, Costa, Soffitta, Krummenacher,
  \& {de Oliveira}"}]{gpd2}
Bellazzini, R., Spandre, G., Minuti, M., {et~al.} 2006, Nuclear Instruments and
  Methods in Physics Research Section A: Accelerators, Spectrometers, Detectors
  and Associated Equipment, 566, 552 ,
  \dodoi{https://doi.org/10.1016/j.nima.2006.07.036}

\bibitem[{Bellazzini {et~al.}(2007)Bellazzini, Spandre, Minuti, Baldini, Brez,
  Latronico, Omodei, Razzano, Massai, Pesce-Rollins, {Sgró}, Costa, Soffitta,
  Sipila, \& Lempinen"}]{gpd3}
---. 2007, Nuclear Instruments and Methods in Physics Research Section A:
  Accelerators, Spectrometers, Detectors and Associated Equipment, 579, 853 ,
  \dodoi{https://doi.org/10.1016/j.nima.2007.05.304}

\bibitem[{Bongiorno {et~al.}(2021)Bongiorno, Kolodziejczak, Kilaru, Eng, Stahl,
  Baumgartner, Thomas, Ranganathan, Ramsey, \& Tucker}]{bongiorno}
Bongiorno, S.~D., Kolodziejczak, J.~J., Kilaru, K., {et~al.} 2021, in Optics
  for EUV, X-Ray, and Gamma-Ray Astronomy X, ed. S.~L. O'Dell, J.~A. Gaskin, \&
  G.~Pareschi, Vol. 11822, International Society for Optics and Photonics
  (SPIE), 189 -- 200, \dodoi{10.1117/12.2594316}

\bibitem[{Costa {et~al.}(2001)Costa, Soffitta, Bellazzini, Brez, Lumb, \&
  Spandre}]{gpd1}
Costa, E., Soffitta, P., Bellazzini, R., {et~al.} 2001, Nature, 411, 662,
  \dodoi{10.1038/35079508}

\bibitem[{{Di Marco} {et~al.}(2022){Di Marco}, Costa, Muleri, Soffitta,
  Fabiani, {La Monaca}, Rankin, Xie, Bachetti, Baldini, Baumgartner,
  Bellazzini, Brez, Castellano, Monte, Lalla, Ferrazzoli, Latronico, Maldera,
  Manfreda, O'Dell, Perri, Pesce-Rollins, Puccetti, Ramsey, Ratheesh, Sgrò,
  Spandre, Tennant, Tobia, Trois, \& Weisskopf}]{DiMarco21}
{Di Marco}, A., Costa, E., Muleri, F., {et~al.} 2022, A weighted analysis to
  improve the X-ray polarization sensitivity of IXPE.
\newblock \doarXiv{2202.01093}

\bibitem[{Feng {et~al.}(2020)Feng, Li, Long, Bellazzini, Costa, Wu, Huang,
  Jiang, Minuti, Wang, Xu, Yang, Baldini, Citraro, Nasimi, Soffitta, Muleri,
  Jung, Yu, Jin, Zeng, An, Brez, Latronico, Sgro, Spandre, \& Pinchera}]{feng}
Feng, H., Li, H., Long, X., {et~al.} 2020, Nature Astronomy, 4, 511,
  \dodoi{10.1038/s41550-020-1088-1}

\bibitem[{Ferrazzoli {et~al.}(2020)Ferrazzoli, Muleri, Lefevre, Morbidini,
  Amici, Brienza, Costa, Monte, {Di Marco}, Persio, Donnarumma, Fabiani, {La
  Monaca}, Loffredo, Maiolo, Maita, Piazzolla, Ramsey, Rankin, Ratheesh,
  Rubini, Sarra, Soffitta, Tobia, \& Xie}]{Ferrazzoli}
Ferrazzoli, R., Muleri, F., Lefevre, C., {et~al.} 2020, Journal of Astronomical
  Telescopes, Instruments, and Systems, 6, 1 ,
  \dodoi{10.1117/1.JATIS.6.4.048002}

\bibitem[{Heitler(1936)}]{Heitler}
Heitler, W. 1936, International Series of Monographs on Physics, Vol.~5, {The
  quantum theory of radiation} (Oxford: Oxford University Press)

\bibitem[{Kislat {et~al.}(2015)Kislat, Clark, Beilicke, \&
  Krawczynski}]{Kislat}
Kislat, F., Clark, B., Beilicke, M., \& Krawczynski, H. 2015, Astroparticle
  Physics, 68, 45, \dodoi{https://doi.org/10.1016/j.astropartphys.2015.02.007}

\bibitem[{{La Monaca} {et~al.}(2021){La Monaca}, Fabiani, Lefevre, Morbidini,
  Piazzolla, Amici, Attinà, Brienza, Costa, Cosimo, {Di Marco}, Persio,
  Evangelista, Ferrazzoli, Loffredo, Muleri, Rankin, Ratheesh, Rubini, Santoli,
  Scalise, Soffitta, Tobia, Trois, Xie, \& Zeiger}]{UV}
{La Monaca}, F., Fabiani, S., Lefevre, C., {et~al.} 2021, in Space Telescopes
  and Instrumentation 2020: Ultraviolet to Gamma Ray, ed. J.-W.~A. den Herder,
  S.~Nikzad, \& K.~Nakazawa, Vol. 11444, International Society for Optics and
  Photonics (SPIE), 1029 -- 1039, \dodoi{10.1117/12.2567000}

\bibitem[{Long {et~al.}(2022)Long, Feng, Li, Zhu, Wu, Huang, Minuti, Jiang,
  Yang, Citraro, Nasimi, Yu, Jin, Zeng, An, Jiang, Costa, Baldini, Bellazzini,
  Brez, Latronico, Sgr{\`{o}}, Spandre, Pinchera, Muleri, \& Soffitta}]{ScoX1}
Long, X., Feng, H., Li, H., {et~al.} 2022, The Astrophysical Journal Letters,
  924, L13, \dodoi{10.3847/2041-8213/ac4673}

\bibitem[{Muleri(2014)}]{Muleri2014}
Muleri, F. 2014, The Astrophysical Journal, 782, 28,
  \dodoi{10.1088/0004-637x/782/1/28}

\bibitem[{Muleri {et~al.}(2016)Muleri, Soffitta, Baldini, Bellazzini, Brez,
  Costa, Lalla, Monte, Evangelista, Latronico, Manfreda, Minuti, Pesce-Rollins,
  Pinchera, Rubini, Sgrò, Spada, \& Spandre}]{Muleri16}
Muleri, F., Soffitta, P., Baldini, L., {et~al.} 2016, in Space Telescopes and
  Instrumentation 2016: Ultraviolet to Gamma Ray, ed. J.-W.~A. den Herder,
  T.~Takahashi, \& M.~Bautz, Vol. 9905, International Society for Optics and
  Photonics (SPIE), 1401 -- 1407.
\newblock \url{https://doi.org/10.1117/12.2233206}

\bibitem[{Muleri {et~al.}(2018)Muleri, Lefevre, Piazzolla, Morbidini, Amici,
  Attina, Centrone, Monte, Cosimo, Persio, Evangelista, Fabiani, Ferrazzoli,
  Loffredo, Maiolo, Maita, Primicino, Rankin, Rubini, Santoli, Soffitta, Tobia,
  Tortosa, \& Trois}]{ICE}
Muleri, F., Lefevre, C., Piazzolla, R., {et~al.} 2018, in Space Telescopes and
  Instrumentation 2018: Ultraviolet to Gamma Ray, ed. J.-W.~A. den Herder,
  S.~Nikzad, \& K.~Nakazawa, Vol. 10699, International Society for Optics and
  Photonics (SPIE), 1312 -- 1322, \dodoi{10.1117/12.2312203}

\bibitem[{Muleri {et~al.}(2021)Muleri, Piazzolla, {Di Marco}, Fabiani, {La
  Monaca}, Lefevre, Morbidini, Rankin, Soffitta, Tobia, Xie, Amici, Attinà,
  Bachetti, Brienza, Centrone, Costa, {Del Monte}, {Di Cosimo}, {Di Persio},
  Evangelista, Ferrazzoli, Loffredo, Perri, Pilia, Puccetti, Ratheesh, Rubini,
  Santoli, Scalise, \& Trois}]{Muleri21}
Muleri, F., Piazzolla, R., {Di Marco}, A., {et~al.} 2021, Astroparticle
  Physics, 102658, \dodoi{https://doi.org/10.1016/j.astropartphys.2021.102658}

\bibitem[{{O'Dell} {et~al.}(2019){O'Dell}, {Attinà}, {Baldini}, {Barbanera},
  {Baumgartner}, {Bellazzini}, {Bladt}, {Bongiorno}, {Brez}, {Cavazzuti},
  {Citraro}, {Costa}, {Deininger}, {Del Monte}, {Dietz}, {Di Lalla},
  {Donnarumma}, {Elsner}, {Fabiani}, {Ferrazzoli}, {Guy}, {Kalinowski},
  {Kaspi}, {Kelley}, {Kolodziejczak}, {Latronico}, {Lefevre}, {Lucchesi},
  {Manfreda}, {Marshall}, {Masciarelli}, {Matt}, {Minuti}, {Muleri}, {Nasimi},
  {Nuti}, {Orsini}, {Osborne}, {Perri}, {Pesce-Rollins}, {Peterson},
  {Pinchera}, {Puccetti}, {Ramsey}, {Ratheesh}, {Romani}, {Santoli},
  {Sciortino}, {Sgrò}, {Smith}, {Spandre}, {Soffitta}, {Tennant}, {Tobia},
  {Trois}, {Wedmore}, {Weisskopf}, {Xie}, {Zanetti}, {Alexander}, {Allen},
  {Amici}, {Antoniak}, {Bonino}, {Borotto}, {Breeding}, {Bygott}, {Caporale},
  {Cardelli}, {Ceccanti}, {Centrone}, {Di Persio}, {Evangelista}, {MacKenzie},
  {Footdale}, {Forsyth}, {Foster}, {Gunji}, {Gurnee}, {Hibbard}, {Johnson},
  {Kelly}, {Kilaru}, {La Monaca}, {Le Roy}, {Loffredo}, {Magazzu}, {Marengo},
  {Marrocchesi}, {Massaro}, {McCracken}, {McEachen}, {Mereu}, {Mitchell},
  {Mitsuishi}, {Morbidini}, {Mosti}, {Negro}, {Oppedisano}, {Pacheco}, {Paggi},
  {Pavelitz}, {Pentz}, {Piazzolla}, {Porter}, {Profeti}, {Ranganathan},
  {Rankin}, {Root}, {Rubini}, {Ruswick}, {Sanchez}, {Scalise}, {Schindhelm},
  {Speegle}, {Tamagawa}, {Tardiola}, {Walden}, {Weddendorf}, \&
  {Welch}}]{ixpe3}
{O'Dell}, S.~L., {Attinà}, P., {Baldini}, L., {et~al.} 2019, in UV, X-Ray, and
  Gamma-Ray Space Instrumentation for Astronomy XXI, ed. O.~H. Siegmund, Vol.
  11118, International Society for Optics and Photonics (SPIE), 248 -- 261,
  \dodoi{10.1117/12.2530646}

\bibitem[{Ramsey {et~al.}(2021)Ramsey, Attina, Baldini, Barbanera, Baumgartner,
  Bellazzini, Bladt, Bongiorno, Brez, Castellano, Carpentiero, Castronuovo,
  Cavalli, Cavazutti, D'Amico, Citraro, Costa, Deininger, D'Alba, Delmonte,
  Dietz, Lalla, Marco, Persio, Donnarumma, Fabiani, Ferrazzoli, Footdale, Head,
  Kalinowski, Kolodziejczak, Latronico, Lefevre, Lorenzi, Lucchesi, Maldera,
  Manfreda, Mangravati, Marshall, Matt, Minuti, Mize, Muleri, Nasimi, Negri,
  Nuti, O'Dell, Orsini, Osborne, Pentz, Pilia, Perri, Pesce-Rollins, Peterson,
  Pinchera, Puccetti, Rankin, Ratheesh, Romani, Sarra, Santoli, Sciortino,
  Schroeder, Sgro, Soffitta, Spandre, Tennant, Tobia, Thomas, Trois, Vimercati,
  Wedmore, Weisskopf, Xie, Zanetti, Alexander, Allen, Amici, Andersen,
  Antonelli, Antoniak, Bachetti, Baggett, Bonino, Boree, Borotto, Breeding,
  Brienza, Byggott, Caporale, Cardelli, Ceccanti, Centrone, Dolan, Evangelista,
  Ferrant, Ferrie, Forsyth, Foster, Garelick, Gunji, Gurnee, Hibbard, Johnson,
  Kelly, Kilaru, Monaca, Roy, Lofredo, Maddox, Magazzu, Marengo, Marrocchesi,
  Massaro, Mauger, McCracken, McEachen, Mereu, Mitchell, Mitsuishi, Morbidini,
  Mosti, Nguyen, Negro, Nitschke, Onizuka, Oppedisano, Pacheco, Paggi, Painter,
  Pavelitz, Piazzolla, Profeti, Ranganathan, Reedy, Root, Rubini, Ruswick,
  Sanchez, Scalise, Seek, Sosdian, Speegle, Tamagawa, Tardiola, Valerie,
  Walden, Weddendorf, \& Welch}]{ixpe5}
Ramsey, B.~D., Attina, P., Baldini, L., {et~al.} 2021, in UV, X-Ray, and
  Gamma-Ray Space Instrumentation for Astronomy XXII, ed. O.~H. Siegmund, Vol.
  11821, International Society for Optics and Photonics (SPIE), 225 -- 236,
  \dodoi{10.1117/12.2595302}

\bibitem[{Rankin {et~al.}(2022)Rankin, Muleri, Tennant, Bachetti, Costa, {Di
  Marco}, Fabiani, {La Monaca}, Soffitta, Tobia, Trois, Xie, Baldini, Lalla,
  Manfreda, O'Dell, Perri, Puccetti, Ramsey, Sgr{\`{o}}, \&
  Weisskopf}]{Rankin2021}
Rankin, J., Muleri, F., Tennant, A.~F., {et~al.} 2022, The Astronomical
  Journal, 163, 39, \dodoi{10.3847/1538-3881/ac397f}

\bibitem[{{Sgrò}(2017)}]{ixpe6}
{Sgrò}, C. 2017, in UV, X-Ray, and Gamma-Ray Space Instrumentation for
  Astronomy XX, ed. O.~H. Siegmund, Vol. 10397, International Society for
  Optics and Photonics (SPIE), 104 -- 110, \dodoi{10.1117/12.2273922}

\bibitem[{Soffitta(2017)}]{ixpe2}
Soffitta, P. 2017, in UV, X-Ray, and Gamma-Ray Space Instrumentation for
  Astronomy XX, ed. O.~H. Siegmund, Vol. 10397, International Society for
  Optics and Photonics (SPIE), 127 -- 135, \dodoi{10.1117/12.2275485}

\bibitem[{Soffitta {et~al.}(2020)Soffitta, Attinà, Baldini, Barbanera,
  Baumgartner, Bellazzini, Bladt, Bongiorno, Brez, Castellano, Carpentiero,
  Castronuovo, Cavalli, Cavazzuti, D'Amico, Citraro, Costa, Deininger, D'Alba,
  Monte, Diets, Lalla, {Di Marco}, Persio, Donnarumma, Elsner, Fabiani,
  Ferrazzoli, Guy, Kalinowski, Kolodziejczak, Latronico, Lefevre, Lorenzi,
  Lucchesi, Maldera, Manfreda, Mangraviti, Marshall, Masciarelli, Matt, Minuti,
  Muleri, Nasimi, Negri, Nuti, Orsini, Osborne, Pilia, Perri, Pesce-Rollins,
  Peterson, Pinchera, Puccetti, Ramsey, Ratheesh, Romani, Sarra, Santoli,
  Sciortino, Sgrò, Smith, Spandre, Tennant, Tobia, Trois, Vimercati, Wedmnore,
  Weisskopf, Xie, Zanetti, Alexander, Allen, Amici, Antonelli, Antoniak,
  Bachetti, Bonino, Borotto, Breeding, Brienza, Bygott, Cardelli, Ceccanti,
  Centrone, Evangelista, Ferrie, Forsyth, Foster, Gurnee, Hibbard, Johnson,
  Kelly, Kilaru, {La Monaca}, Roy, Loffredo, Magazzu', Marengo, Marrocchesi,
  Massaro, Morbidini, McCracken, McEachen, Mereu, Mitchell, Mitsuishi, Mosti,
  Nigro, Nuti, Oppedisano, Pacheco, Paggi, Pavelitz, Pentz, Piazzolla, Porter,
  Profeti, Ranganathan, Rankin, Root, Rubini, Ruswick, Sanchez, Scalise,
  Schindhelm, Speegle, Tamagawa, Tardiola, Walden, Weddendorf, Welch, Head,
  Gray, Mize, O'Dell, Schroeder, Thomas, Bagget, Dolan, Ferrant, Footdale,
  Garelick, Johnson, \& Seek}]{ixpe4}
Soffitta, P., Attinà, P., Baldini, L., {et~al.} 2020, in Space Telescopes and
  Instrumentation 2020: Ultraviolet to Gamma Ray, ed. J.-W.~A. den Herder,
  S.~Nikzad, \& K.~Nakazawa, Vol. 11444, International Society for Optics and
  Photonics (SPIE), 1017 -- 1028, \dodoi{10.1117/12.2567001}

\bibitem[{Soffitta {et~al.}(2021)Soffitta, Baldini, Bellazzini, Costa,
  Latronico, Muleri, Monte, Fabiani, Minuti, Pinchera, Sgro', Spandre, Trois,
  Amici, Andersson, Attina', Bachetti, Barbanera, Borotto, Brez, Brienza,
  Caporale, Cardelli, Carpentiero, Castellano, Castronuovo, Cavalli, Cavazzuti,
  Ceccanti, Centrone, Ciprini, Citraro, D'Amico, D'Alba, Cosimo, Lalla, Marco,
  Persio, Donnarumma, Evangelista, Ferrazzoli, Hayato, Kitaguchi, Monaca,
  Lefevre, Loffredo, Lorenzi, Lucchesi, Magazzu, Maldera, Manfreda, Mangraviti,
  Marengo, Matt, Mereu, Morbidini, Mosti, Nakano, Nasimi, Negri, Nenonen, Nuti,
  Orsini, Perri, Pesce-Rollins, Piazzolla, Pilia, Profeti, Puccetti, Rankin,
  Ratheesh, Rubini, Santoli, Sarra, Scalise, Sciortino, Tamagawa, Tardiola,
  Tobia, Vimercati, \& Xie}]{Soffitta21}
Soffitta, P., Baldini, L., Bellazzini, R., {et~al.} 2021, The Astronomical
  Journal, 162, 208, \dodoi{10.3847/1538-3881/ac19b0}

\bibitem[{Strohmayer \& Kallman(2013)}]{strohmayer}
Strohmayer, T.~E., \& Kallman, T.~R. 2013, The Astrophysical Journal, 773, 103,
  \dodoi{10.1088/0004-637x/773/2/103}

\bibitem[{{Weisskopf} {et~al.}(1976){Weisskopf}, {Cohen}, {Kestenbaum}, {Long},
  {Novick}, \& {Wolff}}]{crab1}
{Weisskopf}, M.~C., {Cohen}, G.~G., {Kestenbaum}, H.~L., {et~al.} 1976, \apjl,
  208, L125, \dodoi{10.1086/182247}

\bibitem[{{Weisskopf} {et~al.}(1978){Weisskopf}, {Silver}, {Kestenbaum},
  {Long}, \& {Novick}}]{crab2}
{Weisskopf}, M.~C., {Silver}, E.~H., {Kestenbaum}, H.~L., {Long}, K.~S., \&
  {Novick}, R. 1978, \apjl, 220, L117, \dodoi{10.1086/182648}

\bibitem[{Weisskopf {et~al.}(2016)Weisskopf, Ramsey, O'Dell, Tennant, Elsner,
  Soffitta, Bellazzini, Costa, Kolodziejczak, Kaspi, Muleri, Marshall, Matt, \&
  Romani}]{ixpe1}
Weisskopf, M.~C., Ramsey, B., O'Dell, S., {et~al.} 2016, in Space Telescopes
  and Instrumentation 2016: Ultraviolet to Gamma Ray, ed. J.-W.~A. den Herder,
  T.~Takahashi, \& M.~Bautz, Vol. 9905, International Society for Optics and
  Photonics (SPIE), 356 -- 365, \dodoi{10.1117/12.2235240}

\bibitem[{Weisskopf {et~al.}(2021)Weisskopf, Soffitta, Baldini, Ramsey, O'Dell,
  Romani, Matt, Deininger, Baumgartner, Bellazzini, Costa, Kolodziejczak,
  Latronico, Marshall, Muleri, Bongiorno, Tennant, Bucciantini, Dovciak, Marin,
  Marscher, Poutanen, Slane, Turolla, Kalinowski, Marco, Fabiani, Minuti,
  Monaca, Pinchera, Rankin, Sgro', Trois, Xie, Alexander, Allen, Amici,
  Andersen, Antonelli, Antoniak, Attina', Barbanera, Bachetti, Baggett, Bladt,
  Brez, Bonino, Boree, Borotto, Breeding, Brienza, Bygott, Caporale, Cardelli,
  Carpentiero, Castellano, Castronuovo, Cavalli, Cavazzuti, Ceccanti, Centrone,
  Citraro, Amico, D'Alba, Gesu, Monte, Dietz, Lalla, Persio, Dolan, Donnarumma,
  Evangelista, Ferrant, Ferrazzoli, Ferrie, Footdale, Forsyth, Foster,
  Garelick, Gunji, Gurnee, Head, Hibbard, Johnson, Kelly, Kilaru, Lefevre, Roy,
  Loffredo, Lorenzi, Lucchesi, Maddox, Magazzu, Maldera, Manfreda, Mangraviti,
  Marengo, Marrocchesi, Massaro, Mauger, McCracken, McEachen, Mize, Mereu,
  Mitchell, Mitsuishi, Morbidini, Mosti, Nasimi, Negri, Negro, Nguyen,
  Nitschke, Nuti, Onizuka, Oppedisano, Orsini, Osborne, Pacheco, Paggi,
  Painter, Pavelitz, Pentz, Piazzolla, Perri, Pesce-Rollins, Peterson, Pilia,
  Profeti, Puccetti, Ranganathan, Ratheesh, Reedy, Root, Rubini, Ruswick,
  Sanchez, Sarra, Santoli, Scalise, Sciortino, Schroeder, Seek, Sosdian,
  Spandre, Speegle, Tamagawa, Tardiola, Tobia, Thomas, Valerie, Vimercati,
  Walden, Weddendorf, Wedmore, Welch, Zanetti, \& Zanetti}]{weisskopf2021}
Weisskopf, M.~C., Soffitta, P., Baldini, L., {et~al.} 2021, The Imaging X-Ray
  Polarimetry Explorer (IXPE): Pre-Launch.
\newblock \doarXiv{2112.01269}

\end{thebibliography}
\bibliographystyle{aasjournal}



\end{document}